\documentclass[superscriptaddress, nofootinbib]{revtex4}[11pt]
\pdfoutput=1

\usepackage{graphicx}
\graphicspath{{figs/}}
\usepackage {amsmath, amssymb, rotating}
\usepackage{hyperref}
\usepackage[usenames,dvipsnames]{color}
\usepackage{mdwlist} 
\usepackage{slashed}

\makeatletter
\def\@seccntformat#1{\csname the#1\endcsname.\quad} 

\renewcommand\section{\@startsection
  {section}{1}{0mm}
  {-\baselineskip}
  {0.5\baselineskip}
  {\normalfont\normalsize\bf}}

\renewcommand\subsection{\@startsection
  {subsection}{2}{0mm}
  {-\baselineskip}
  {0.5\baselineskip}
  {\normalfont\normalsize\bf}}

\renewcommand\paragraph{\@startsection
  {paragraph}{4}{\z@}
  {1.25ex \@plus1ex \@minus.2ex}
  {-1em}
  {\normalfont\normalsize\bfseries}}  
\makeatother

\def\simgt{\mathrel{\lower2.5pt\vbox{\lineskip=0pt\baselineskip=0pt
           \hbox{$>$}\hbox{$\sim$}}}}
\def\simlt{\mathrel{\lower2.5pt\vbox{\lineskip=0pt\baselineskip=0pt
           \hbox{$<$}\hbox{$\sim$}}}}

\newcommand{\mymatrix}[1]{\begin{pmatrix} #1 \end{pmatrix}}

\def\A{A}

\newcommand{\amre}{a_\text{mre}}
\newcommand{\Hmre}{H_\text{mre}}

\begin{document}

\title{Vector Dark Matter from Inflationary Fluctuations}

\author{Peter W. Graham}
\affiliation{Stanford Institute for Theoretical Physics, Department of Physics, Stanford University, Stanford, CA 94305}

\author{Jeremy 
 Mardon}
\thanks{jmardon@stanford.edu}
\affiliation{Stanford Institute for Theoretical Physics, Department of Physics, Stanford University, Stanford, CA 94305}

\author{Surjeet 
Rajendran}
\affiliation{Berkeley Center for Theoretical Physics, Department of Physics, University of California, Berkeley, CA 94720}

\begin{abstract}
  \vskip 3pt \noindent 
We calculate the production of a massive vector boson by quantum fluctuations during inflation.  This gives a novel dark-matter production mechanism quite distinct from misalignment or thermal production.
While scalars and tensors are typically produced with a nearly scale-invariant spectrum, surprisingly the vector is produced with a power spectrum peaked at intermediate wavelengths.  Thus dangerous, long-wavelength, isocurvature perturbations are suppressed.  Further, at long wavelengths the vector inherits the usual adiabatic, nearly scale-invariant perturbations of the inflaton, allowing it to be a good dark matter candidate.  The final abundance can be calculated precisely from the mass and the Hubble scale of inflation, $H_I$.  Saturating the dark matter abundance we find a prediction for the mass $m \approx 10^{-5}\,$eV$\times(10^{14}\,$GeV$/H_I)^4$.  High-scale inflation, potentially observable in the CMB, motivates an exciting mass range for recently proposed direct detection experiments for hidden photon dark matter.  Such experiments may be able to reconstruct the distinctive, peaked power spectrum, verifying that the dark matter was produced by quantum fluctuations during inflation and providing a direct measurement of the scale of inflation.  Thus a detection would not only be the discovery of dark matter, it would also provide an unexpected probe of inflation itself.
\end{abstract}

\maketitle


\tableofcontents


\section{Introduction}
\label{sec:intro}

Inflation \cite{Guth:1980zm, Linde:1981mu} is a compelling model of the earliest moments of the universe. It addresses many theoretical puzzles such as the observed large scale homogeneity and isotropy of the universe. Simultaneously, through quantum mechanical perturbations, it seeds density inhomogeneities that can explain the origin of structure in the universe \cite{Mukhanov:1990me}. These inhomogeneities are imprinted on the cosmic microwave background (CMB), and where probed, their  spectrum matches with those of the predictions of inflation. This remarkable agreement  is often regarded as the best observational evidence for the inflationary paradigm. 

 Cosmological measurements show that the observed growth of structure in our universe requires the existence of a new particle, namely, the dark matter. Conventionally, it is assumed that the origins of dark matter are decoupled from the mechanics of inflation -- candidates such as Weakly Interacting Massive Particles (WIMPs) are assumed to arise from thermal processes that occur after inflation reheats the universe \cite{Kamionkowski:1997zb} while ultra-light scalars such as axions acquire a cosmological abundance as a virtue of initial conditions \cite{Dine:1982ah, Preskill:1982cy}.  While it is reasonable that these two sectors are decoupled, it is tempting to ask if inflation, a theory that so beautifully explains the origins of structure in the universe, could also simultaneously be the source of the dark matter that is essential for the growth of this structure. 
 
In fact, inflation contains a natural mechanism to generate a cosmological abundance of particles.
The transition from the the early inflationary era to the later radiation- or matter-dominated era populates field modes that were initially in the vacuum state \cite{Ford:1986sy, Lyth:1996yj}.
For example, inflation is expected to produce a nearly scale invariant spectrum of gravitational waves~\cite{Mukhanov:1990me}, whose discovery would be regarded as proof of the inflationary paradigm. 
Similarly, a scalar field whose mass is less than the Hubble scale during inflation will generically be coherently populated~\cite{Mukhanov:1990me} with a scale invariant spectrum. 
This could produce an energy density equal to that of cosmic dark matter for a sufficiently light scalar (such as an axion). However, the power spectrum of such a field would contain isocurvature perturbations that do not match CMB observations. 
This rules out this production mechanism for these dark matter candidates, instead leading to constraints on such scenarios in simple models \cite{Fox:2004kb}. 

Given these known results for scalars and tensors, we consider the inflationary production of ultra-light vector bosons {\it i.e.} vector bosons with a small ($\ll 100$ GeV), but non-zero mass. Such vector bosons may arise naturally in frameworks of physics beyond the standard model \cite{Arvanitaki:2009hb}. Further,  much like axions, their interactions with the standard model can be naturally very small, preventing their thermalization with the standard model plasma in the early universe  \cite{Nelson:2011sf, Arias:2012az}. Consequently, an abundance of such particles produced during inflation can be cosmologically interesting. We show that ultra-light vector bosons are produced by inflation.  While they are initially produced with a spectrum similar to that of scalars and tensors,  importantly  their spectrum evolves differently during cosmological expansion. As a result,  inflationary production of these vector bosons can account for {\it all} of the observed dark matter density. Further, unlike the case for scalars, we show that the spectrum of density inhomogeneities produced by this mechanism matches with those observed in the CMB. This can be accomplished without any additional terms in the Lagrangian (such as couplings to the standard model) -- all that is necessary is the existence of a massive vector boson. While the spectrum is automatically consistent with the CMB, to match the observed dark matter abundance,  the mass of the vector boson is a function of the inflationary scale. Thus the existence of an ultra-light massive vector boson and inflation is sufficient to produce the observed properties of dark matter in the universe. 

Naively, this difference in the spectrum of scalars (or tensors) and vectors would seem to be a surprise. After all, the perturbations responsible for the production of these particles is entirely the result of gravitational dynamics during inflation. 
It might seem that the equivalence principle would lead to identical spectra for all these bosons. 
Indeed, for modes that are inside the horizon, the energy density in scalars and vectors behaves identically. 
However, the evolution of the energy density in super horizon modes of the vector boson is different from those of scalars. 
As a result, the final spectrum is different. This is thus consistent with the equivalence principle, since the equivalence principle, being a local statement,  only guarantees identical evolution of sub-horizon fluctuations. The differences in the evolution of super horizon modes is precisely the reason for the well known fact that  while inflation copiously produces massless scalars and tensors (gravitational waves), it does not source massless vectors. 

The cosmological evolution of such ultra-light massive vector bosons have been considered before~\cite{Nelson:2011sf, Arias:2012az, Dimopoulos:2006ms}, and the results of~\cite{Arias:2012az} on the evolution of transverse modes and of~\cite{Dimopoulos:2006ms} on the evolution of super-horizon longitudinal modes agree with our calculations. However,~\cite{Arias:2012az} was focussed on preserving a dark matter abundance of these vector  bosons produced by the misalignment mechanism, wherein the field is generically assumed to have a non-zero initial value prior to inflation. Owing to the red-shift in energy density,  \cite{Arias:2012az} concluded that preservation of the dark matter density through an initial misalignment would require the inclusion of additional terms in the Lagrangian with a fine tuned choice of coefficients. Further, they did not focus on the fact that a massive vector boson has three degrees of freedom - two transverse modes and a longitudinal mode. While the production of the transverse modes is suppressed during inflation (they are approximately conformal, much like massless vectors), we find that the longitudinal mode is copiously produced. This production is central to our result. 
The production of the longitudinal mode and its super-horizon evolution was previously considered in~\cite{Dimopoulos:2006ms}. However, that paper was focussed on reheating the universe through the direct decays of a massive vector boson, and did not consider the late time abundance. 
Since in that scenario the vector was directly responsible for the CMB power spectrum, it needed to be scale invariant, and hence~\cite{Dimopoulos:2006ms} found that such scenarios are constrained. 

The focus of our work is different -- the longitudinal modes of a massive vector boson sourced by inflation becomes the dark matter of the universe. Over the wavelengths of perturbations observed in the CMB, the dominant density perturbations of the vector field are inherited from the inflaton in the same manner as they are for the radiation field, 
thus matching observations of the CMB. 
For this to be true, unlike \cite{Arias:2012az, Dimopoulos:2006ms}, we do not need any additional terms -- as long as there is a vector boson with the correct mass, inflationary production generates a cosmological abundance for it that is consistent with observation. 
These results~\cite{Fermilab:2014} demonstrate a novel, purely gravitational production mechanism for dark-matter, which occurs automatically during inflation, and appears to be unique to light vector fields.\footnote{Although for a related production mechanism for extremely heavy dark matter see~\cite{Chung:1998zb}.}

We begin in section \ref{Sec:Exec} with an executive overview of the mechanism and results. 
In section \ref{sec:vector-in-FRW} we derive the main result, beginning with the theory of a vector in the expanding universe, following its production and subsequent evolution, and finally determining the final relic abundance.
The correct imprinting of adiabatic density fluctuations is shown in section \ref{sec:fluctuations}. 
We discuss the phenomenology and direct detection prospects of such dark matter in section \ref{Sec:Pheno}. 
Finally we conclude in section \ref{Sec:Conclusions}.

\section{Executive Summary}
\label{Sec:Exec}

We calculate the production of a massive vector boson by quantum fluctuations during inflation.
The \emph{longitudinal} modes are produced with a non-scale-invariant power spectrum, while the transverse modes are suppressed.
If the vector is cosmologically stable, the abundance produced ends up as cold relic matter in the late universe, in the form of a coherent oscillating condensate of the field. 
This novel production mechanism has interesting consequences that we explore.

\paragraph*{Spectrum}
For commonly considered light fields, the spectrum of fluctuations produced by inflation is flat (or nearly flat) over a large range of wavelengths. This is true for the inflaton itself, for the graviton, and for any canonically coupled light scalar field.  
However, we find that the longitudinal modes of a canonically coupled massive vector\footnote{We assume the mass of the vector is ``on'' during and after inflation, such as in the case of a Stueckelberg (i.e. fundamental) mass. Our results do not apply, for example, if the mass is only generated after reheating by the Higgs mechanism.} are produced with a peaked spectrum (see Fig.~\ref{fig:density-power-spectra}).  
The peak occurs at an intermediate wavelength, much smaller than the size of the observable universe but much larger than the usual short distance cutoff of the inflationary spectra.  The power is greatly suppressed at both large and small wavelengths. 

The fact that the spectrum is not scale-invariant arises from an interesting difference in the behavior of a scalar field and a vector field in the early universe.  
The energy density in non-relativistic modes of a scalar field is frozen when Hubble is greater than its mass. 
In contrast, the energy in a vector redshifts as $1/a^2$ in this regime.  This affects the long-wavelength modes the most (see Fig.~\ref{fig:mode-evolution-2}), suppressing power on large scales. 
One important consequence of this is that the misalignment mechanism is ineffective at producing a late-time abundance of a vector.
Unlike for a scalar such as the QCD axion, energy density stored in a homogeneous vector field damps away while $H > m$, making any relic abundance produced in this way negligible\footnote{This applies to both transverse and longitudinal modes, since there is no distinction between them in the zero-momentum limit.}$^,$\footnote{The failure of the misalignment mechanism was noted in Ref.~\cite{Arias:2012az}. There, a large coupling of the vector to the curvature $\mathcal R$ was introduced to restore the possibility of misalignment production. However, such a coupling would generically be expected to introduce a quadratic divergence for the vector mass, and so we do not consider it here.}. 
The difference from a scalar field arises from the fact that the metric (and hence scale factor) is inherently part of the norm of the vector (simply relating $A_\mu$ to $A^\mu$ brings in the scale factor).

\paragraph*{Successful density fluctuations} This production mechanism is perhaps most interesting as a way to generate the measured dark matter abundance.  
With regard to this, a second important consequence of the long-wavelength suppression is the absence of isocurvature modes on cosmological scales, which would be visible in the CMB (and are ruled out by observations).  
For a scalar (such an axion), such isocurvature modes are dangerous and typically rule out large parts of parameter space.   
However, the vector naturally avoids these observational constraints.

Of course, to be a good dark matter candidate the vector must also have the standard scale-invariant, adiabatic, $\delta \rho/\rho\!\sim\!10^{-5}$ fluctuations on cosmological scales, which are understood to be the imprint of earlier inflaton fluctuations.  
Luckily, these fluctuations are automatically imprinted onto the density of the vector.  
This can be understood intuitively from the ``Separate Universes'' approach. Since the observed adiabatic fluctuations occur at wavelengths far larger than the scale of the dominant vector fluctuations, these wavelengths are far outside the horizon while all the relevant vector dynamics is occurring.  
Each patch of the universe can therefore be treated as a separate homogeneous universe, with the local inflaton fluctuation affecting it only as an overall offset in the local clock or scale factor.
Because this is just a change to the overall clock, it affects massive vector dark matter in the same way as it would affect any type of dark matter (e.g.~a thermally produced relic particle).  So, just as in a normal scenario, the dark matter density will fluctuate along with the densities of every other component of the universe.
This guarantees the correct adiabatic fluctuation spectrum.
Thus inflationary fluctuations of a light vector provide a successful, and novel, dark matter production mechanism.

\paragraph*{Relic abundance} In order to make up the measured dark matter abundance we find a condition on the mass of the vector and the Hubble scale of inflation, given by
\begin{equation}
\Omega_A = \Omega_{\rm cdm} \times \sqrt{\frac{m}{6\times10^{-6}\,\text{eV}}} \, \bigg(\frac{H_I}{10^{14}\,\text{GeV}} \bigg)^2 \, .
\end{equation}
The current bound on the scale of inflation, $H_I \simlt 10^{14}\, \text{GeV}$, therefore places a lower limit on the vector mass required for this mechanism to generate the entire dark matter abundance, $m\simgt 10^{-5} \,\text{eV}$. 
A lighter vector could nonetheless make up an interesting sub-component of the dark matter abundance.
Requiring $m < H$ during inflation allows the mechanism to work in principle for masses as large as $\sim\!10^8\,\text{GeV}$.

\paragraph*{Direct detection phenomenology} A new vector field will generically have, at some level, a kinetic mixing with the photon. 
This would cause the vector field to couple weakly to charged Standard Model particles, enabling detection of the dark matter with various proposed or existing experimental setups operating in different mass ranges~\cite{Arias:2012az, Horns:2012jf, Dobrich:2014kda, Chaudhuri:2014dla, An:2014twa}. 
In particular, a relatively high scale of inflation would put the vector in the optimal mass range for the experiment recently proposed in~\cite{Chaudhuri:2014dla}.
This experiment would take advantage of the fact that the vector field is coherently oscillating at a fixed frequency (set by its mass), to cover a wide range of possible vector masses with high sensitivity (see Fig.~\ref{fig:direct-detection}).

An interesting feature of the inflationary production mechanism is the large power in fluctuations near the peak of the spectrum, at a comoving scale
\begin{equation}
L_{\rm comoving} \sim 10^{10}\,\text{km} \times \sqrt{\frac{10^{-5}\,\text{eV}}{m}} \, .
\label{eq:comoving-scale}
\end{equation}
On these scales, dark matter would therefore be expected to clump and form self-bound substructures (these would form early, resulting in a smaller physical size than indicated in Eq.~\ref{eq:comoving-scale}). 
Since the earth is moving at around $10^{10}\,\text{km/year}$, direct detection experiments would be able to map out much of this interesting structure on experimental timescales, assuming the structure survives gravitational disruption in the galaxy.
By both measuring the vector's mass and inferring its original power spectrum would allow a dramatic confirmation of the inflationary origin of the dark matter, as well as giving an entirely new and powerful way to probe inflation itself.

\section{Relic abundance of a vector from inflationary fluctuations}
\label{sec:relic-abundance}
In this section, we show that quantum fluctuations during inflation can produce a  calculable relic abundance of massive vectors that depends only on the Hubble scale $H_I$ during inflation and the mass $m$ of the massive vector boson. The computation is similar to  the calculation of  inflationary production of scalars and tensors. The massive vector field is decomposed into spatial Fourier modes, which are initially sub-horizon and evolve to become super-horizon modes during inflation. This evolution populates these Fourier modes. Subsequently, the modes re-enter the horizon during radiation domination. To compute the abundance and spectrum of the particle today, it is necessary to track the evolution of the mode on both super and sub-horizon scales. 

The key point  in this calculation is that a massive vector boson has three physical degrees of freedom - two transverse modes and one longitudinal mode. The transverse modes, being approximately conformally invariant, are not efficiently produced by inflation. The longitudinal mode suffers no such suppression and is efficiently produced. This is one of the central results of this section. To show this, we decompose a massive vector into transverse and longitudinal modes in sub-section \ref{sec:vector-in-FRW} and identify the behavior of these modes under various regimes. In particular, we identify a regime where the longitudinal modes are relativistic and evolve like massless scalars. Using this result, in sub-section \ref{sec:inflationary-fluctuations}, we show that much like massless scalars, inflation will also produce these longitudinal modes. In sub-section \ref{sec:evolution}, we track the evolution of the produced modes as they transform from being super-horizon to sub-horizon modes. Using the results of sub-section \ref{sec:vector-in-FRW}, we show that the energy density in non-relativistic, Hubble-damped, super-horizon modes of this field redshifts with the expansion of the universe. This is in contrast to massless scalars and results in the production of a non-scale-invariant spectrum for the massive vector boson. Finally, in sub-section \ref{sec:final-abundance} we calculate the relic abundance of the massive vector as a function of its mass and the Hubble scale during inflation. Using this spectrum as the input, we show in section \ref{sec:fluctuations} that the adiabatic perturbations of the inflaton are also imprinted on the spectrum of the massive vector boson, enabling it to be a good dark matter candidate.

\subsection{A massive vector in the expanding universe}
\label{sec:vector-in-FRW}
We consider the following action for a massive vector 

\begin{equation}
S = \int  \sqrt{|g|} \, d^3 x \, dt \Big( -\frac{1}{4} g^{\mu \kappa} g^{\nu \lambda} F_{\mu \nu} F_{\kappa \lambda} - \frac{1}{2} m^2 g^{\mu \nu} A_\mu A_\nu  \Big) \, .
\label{eq:fundamental-action}
\end{equation}
where $F_{\mu \nu}$ is the field strength of the massive vector boson $A_{\mu}$. Here, the mass is assumed to be a Stueckelberg mass. For an Abelian gauge boson, such a mass term is simply a parameter in the Lagrangian and is technically natural. Our analysis will also apply to cases where such a mass term is generated dynamically at a scale higher than the inflationary scale. All that is relevant for this analysis is that  \eqref{eq:fundamental-action} is the correct effective action for the massive vector field at all times during and after inflation. This field has three degrees of freedom - two transverse modes and one longitudinal mode. Note that the longitudinal mode still exists in the spectrum in the limit $m\to0$. In this limit, it is decoupled from other degrees of freedom but its coupling to gravity is unsuppressed. This is evident from the Goldstone boson equivalence theorem since this limit can be obtained by taking the  $v\to\infty$, $g\to0$ limit of the Higgs mechanism. The coupling of the longitudinal mode to gravity is unsuppressed and as we will see below, it is not conformally invariant. Hence, it can be efficiently sourced by inflation, much like scalar or tensor fields. 

To analyze these effects, we start with the equations of motion of the massive vector boson field in the background of an expanding FRW universe. For the production mechanism, it is sufficient to focus on homogenous and isotropic cosmologies as these will generate an initial spectrum for the field. The effects of inflationary density fluctuations on this spectrum are phenomenologically important and are calculated in   section \ref{sec:fluctuations}. The metric of an FRW expanding universe is
\begin{align}
g_{\mu \nu} = \mymatrix{ -1 & \\  & a(t)^2 \delta_{i j}} \, .
\label{eq:FRW-metric}
\end{align}
where $a(t)$ is the scale factor.  In components, the action is then
\begin{align}
S =  &\int a^3 d^3 x \, dt \,
\frac{1}{2}  \bigg( \frac{1}{a^2} \big| \partial_t \vec A - \vec \nabla A_t \big|^2 -\frac{1}{a^4} \big| \vec \nabla \times \vec A\big|^2 + m^2 A_t^2 -\frac{1}{a^2} m^2 \big|\vec A \big|^2   \bigg) 
\label{eq:action-in-components}
\end{align}
where $\vec A = A_i$ (as opposed to than $A^i$). Note that $A_t$ does not have a kinetic term and it is thus a non-dynamical auxiliary field.  Fourier transforming allows us to complete the square and decouple it from the physical propagating modes:
\begin{align}
S  =  &\int \frac{a^3 d^3 k \, d t}{(2 \pi)^3 }  \, \bigg[ 
\overbrace{\frac{1}{2} \Big(\frac{k^2}{a^2} + m^2 \Big)  \bigg| \A_t - \frac{i \vec k \cdot \partial_t \vec \A}{k^2 + a^2 m^2} \bigg|^2
\vphantom{\frac{\big|^2}{k^2}}}^\text{decoupled auxiliary field}
\nonumber \\
&\hspace{5cm}
+  
\underbrace{\frac{1}{2 a^2}  \Big(
\big| \partial_t \vec \A \big|^2 - \frac{\big| \vec k \cdot \partial_t \vec \A \big|^2}{k^2 + a^2 m^2} -\frac{1}{a^2} \big| \vec k \times \vec \A\big|^2 - m^2 \big|\vec \A \big|^2   \Big)}_\text{physical modes}
 \Bigg] \, .
\label{eq:action-aux-plus-physical}
\end{align}
where $k$ is the co-moving momentum. 
Note that since $\A(\vec x, t)$ is real, its spatial Fourier transform obeys $\A(\vec k, t) = \A(-\vec k, t)^*$.
We can separate $\vec \A$ into transverse and longitudinal components $\vec \A_T$ and $\A_L$, where $\vec k \cdot \vec \A = k \A_L$ and $\vec k \cdot \vec \A_T = 0$.\footnote{Note that completing the square to give Eq.~\ref{eq:action-aux-plus-physical} is equivalent to setting $\A_t = i \vec k \cdot \partial_t \vec \A/(k^2 + a^2 m^2) \propto \A_L$ in the classical equations of motion. Assuming $\A_t = 0$ therefore amounts to throwing out the longitudinal hidden-photon modes.}
 The action separates into an action for the transverse modes, 
\begin{gather}
S_{\rm Trans} = \int \frac{a^3 d^3 k \, d t}{(2 \pi)^3} \,  \frac{1}{2 a^2}
\bigg( \big| \partial_t \vec \A_T \big|^2 - \Big(\frac{k^2}{a^2} + m^2 \Big) \big| \vec \A_T \big|^2  \bigg) \, ,
\label{eq:action-trans}
\end{gather}
and one for the longitudinal modes,
\begin{gather}
S_{\rm Long} = \int \frac{a^3 d^3 k \, d t}{(2 \pi)^3} \,  \frac{1}{2 a^2} 
\Big( \frac{a^2 m^2}{k^2 + a^2 m^2} \big| \partial_t \A_L \big|^2 - m^2 \big| \A_L \big|^2   \Big) \, .
\label{eq:action-long}
\end{gather}

Switching to the conformal time coordinate $d t = a d \eta$, the action for the transverse modes can be written
\begin{gather}
S_{\rm Trans} = \int \frac{d^3 k \, d \eta}{(2 \pi)^3} \,  \frac{1}{2}
\bigg( \big| \partial_\eta \vec \A_T \big|^2 - \Big(k^2 + a^2 m^2 \Big) \big| \vec \A_T \big|^2  \bigg) \, .
\label{eq:action-trans}
\end{gather}
Here the almost complete disappearance of $a$ displays the well-known conformal symmetry of the transverse modes. It is broken only by the small mass term.  This means that, unlike the longitudinal modes, production of transverse modes by inflationary fluctuations is negligible.  We focus on the longitudinal modes for the remainder of the paper.

The action for the longitudinal modes, Eq.~(\ref{eq:action-long}), has an unusual form. However, field redefinitions allow us to recover more familiar forms in the both the high-energy and late-time limits. At early times, when $a \, m \ll k$ ({\it i.e.} the relativistic limit, where the physical momentum of the mode is much bigger than its mass), the $a^2 m^2$ term can be dropped from the denominator in Eq.~(\ref{eq:action-long}). With a simple field redefinition ,
\begin{equation}
\pi (\vec k, t) \equiv \frac{m}{k} \A_L (\vec k, t) \, ,
\label{eq:NGB-field-redef}
\end{equation}
the action becomes familiar:
\begin{align}
S_{\rm Long} \xrightarrow{a m \ll k} 
\int \frac{a^3 d^3 k}{(2 \pi)^3} \, dt
\frac{1}{2} \Big(| \partial_t \pi |^2 - \frac{k^2}{a^2} | \pi |^2   \Big)   
= \int a^3 d^3 x \, dt \frac{1}{2} \Big( ( \partial_t \pi )^2 - \frac{1}{a^2} \big| \vec \nabla \pi \big|^2   \Big) \, .
\label{eq:relativistic-action}
\end{align}
We see that the field $\pi$ has the usual action of a real massless scalar field in the expanding universe. As expected from the Goldstone Boson Equivalence Theorem, in the highly relativistic limit the longitudinal mode is equivalent to a massless (pseudo)scalar. As $m \to 0$, the interactions of $\pi$ with gravity are unaffected while it decouples from other sectors of the theory. Hence, this mode can be produced gravitationally even when $m \to 0$.

\begin{figure}[t]
\begin{center}
\includegraphics[width=0.9\textwidth]{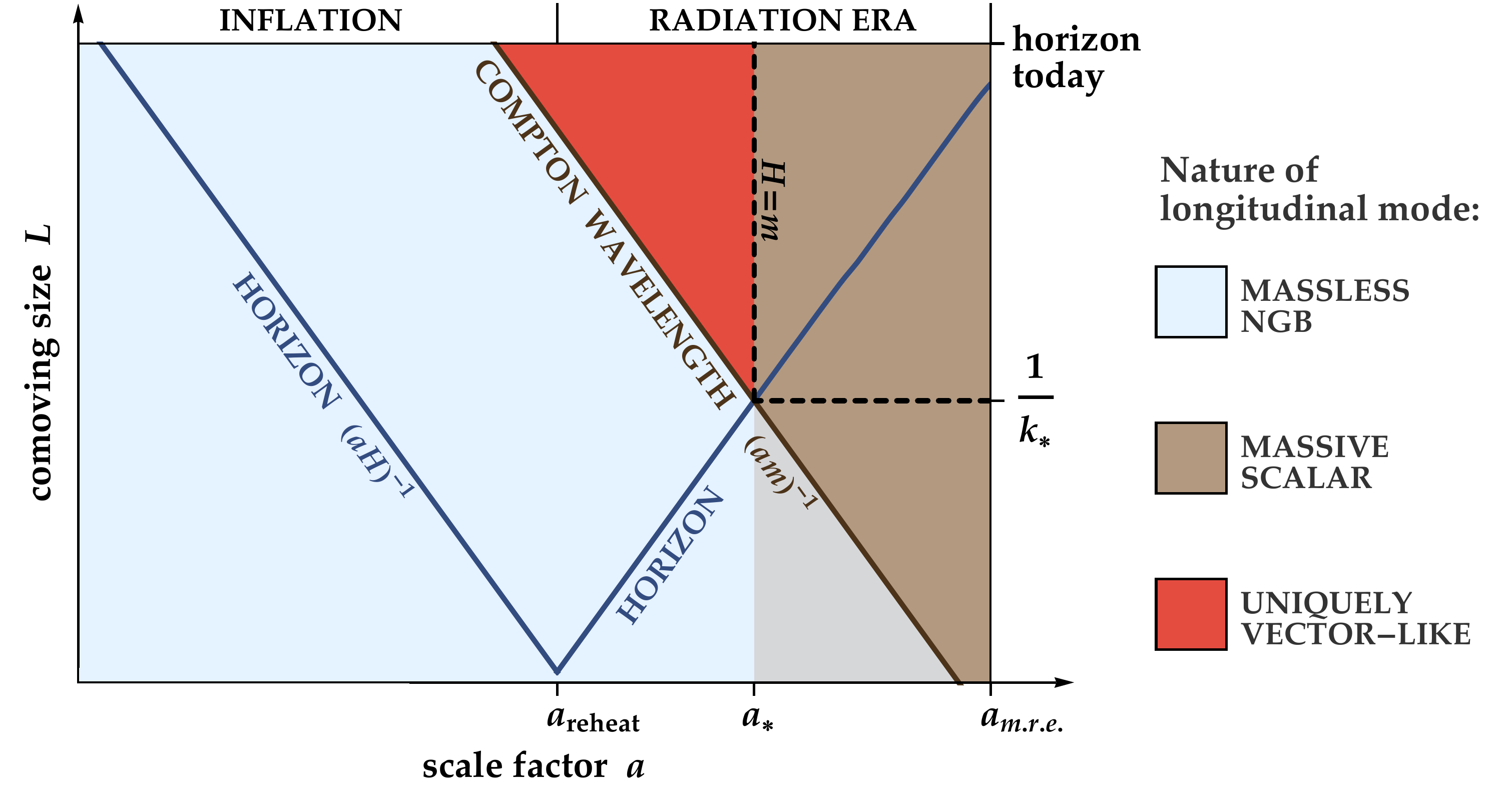}
\caption{Cosmological evolution of length scales, and nature of the longitudinal mode in different regimes.
The line labelled ``horizon'' shows the comoving horizon size $1/a H$, which shrinks during inflation and grows after reheating. 
The line labelled ``Compton wavelength'' shows the comoving Compton wavelength of the vector, $1/a m$.
Modes of the vector field maintain fixed comoving wavevector $k$, evolving along straight lines from left to right.
In the pale blue shaded region, where modes are relativistic ($m\ll k/a$), the longitudinal mode behaves identically to a massless Nambu-Goldstone boson.
In the pale brown shaded region, where Hubble damping is not important ($H \ll m$), the longitudinal mode behaves identically to a free massive scalar.
In the red triangle between these regions, the longitudinal mode has a new behaviour unlike any scalar. 
Modes crossing the tip of this region reenter the horizon just as they become non-relativistic -- their wavevector defines the special scale $k_*$.
}
\label{fig:mode-evolution-1}
\end{center}
\end{figure}

At late times, when the Hubble parameter $H$ is small, the expansion of the universe  has only a small effect on the field. It is then useful to make the field redefinition which makes the longitudinal mode's kinetic term canonical:
\begin{equation}
\phi(\vec k, t) \equiv \frac{m}{\sqrt{k^2 + a^2 m^2}} \A_L(\vec k, t) \, .
\label{eq:non-rel-field-redef}
\end{equation}

The $a$-dependence of this redefinition introduces a term of order $H^2$ into the potential:
\begin{align}
S_{\rm Long}
&= \int \frac{a^3 d^3 k}{(2 \pi)^3} \, dt
\frac{1}{2} \Big(\big| \partial_t  \phi + \frac{a^2 m^2}{k^2 + a^2 m^2}H  \phi \big|^2 - \Big(m^2 + \frac{k^2}{a^2}\Big) |  \phi |^2   \Big)    \\
&= \int \frac{a^3 d^3 k}{(2 \pi)^3} \, dt \frac{1}{2} \Big(| \partial_t  \phi |^2 - \Big(m^2 + \frac{k^2}{a^2}  \Big) |  \phi |^2 +  \mathcal O (H^2) |  \phi |^2    \Big)
\end{align}
When $m^2 \ll H^2$, the Hubble term can be dropped (in fact, keeping track of the exact coefficient shows that it can also be dropped at very early times when $k^2 \gg a^2 m H$, but we have already dealt with this regime). 
The action then reduces to the usual action for a real massive scalar in the expanding universe 
\begin{align}
S_{\rm Long} \xrightarrow{H \ll m}
\int a^3 d^3 x \, dt \frac{1}{2} \Big( (\partial_t \phi )^2  - \frac{1}{a^2} | \vec \nabla \phi |^2 - m^2 \phi^2 \Big) \, .
\label{eq:late-time-effective-action}
\end{align}

In figure~\ref{fig:mode-evolution-1} we show the different regimes for the behaviour of the longitudinal modes through the expansion of the universe.
The pale blue region shows the relativistic regime, where longitudinal modes behave the same way as modes of a free massless scalar. The late-time behavior is shown in the brown region where the modes behave as the modes of a free massive scalar.  In this regime, the longitudinal modes are sub-horizon and as we will see, the energy density in these modes red-shifts as $a^{-3}$. This is of course identical to the red-shift of the energy density of matter and this behavior is expected from the equivalence principle. In the red shaded triangle, neither of these simplifications apply, and the longitudinal modes have a uniquely vector-like behavior which must be determined from its equations of motion. In this regime, the energy density in these modes red-shifts differently than the corresponding modes of scalars. However, this behavior does not violate the equivalence principle as these longitudinal modes are super-horizon and the equivalence principle is only a statement about local interactions of fields in a gravitational background.   

In a universe that undergoes inflation and subsequently reheats, all of these regimes of behavior are important. During the inflationary phase, relativistic, sub-horizon modes get stretched, become horizon size and eventually exit the horizon. As we will see in sub section \ref{sec:inflationary-fluctuations}, this process leads to particle production, populating the mode. After the end of inflation, the universe reheats and the Hubble scale decreases. These modes re-enter the horizon, subsequently evolving to become  sub-horizon, non-relativistic modes, during which they will go through the red-shaded region. As we will see, the red-shifting of the energy density of these longitudinal modes in the red-shaded region is crucial in suppressing dangerous, long-wavelength isocurvature power in the spectrum of the massive vector boson.

The energy density in the longitudinal modes is found using the canonical stress-energy tensor of the field which yields $\rho = 2 g^{0 0} \frac{\partial \mathcal L}{\partial g^{0 0}} - \mathcal L$.  Fourier transforming the action of Eq.~\ref{eq:action-long} back into coordinate space, and restoring $g^{0 0}$, allows the lagrangian to be written in the form
\begin{gather}
\mathcal L = \frac{1}{2 a^2}  \Big( -g^{0 0} \partial_t A_L \frac{a^2 m^2}{a^2 m^2 - \nabla^2} \partial_t A_L - m^2 A_L^2   \Big) \, ,
\end{gather}
from which the energy density follows,
\begin{align}
\rho &= \frac{1}{2 a^2} \Big(
 \partial_t A_L \frac{a^2m^2}{a^2 m^2 - \nabla^2} \partial_t A_L + m^2 A_L^2   \Big) \, .
\end{align}

The Laplacian in the denominator should be understood in terms of its action on Fourier modes, and this non-local form arises because we integrated out the time-component of the vector field.
In a region much longer than the typical wavelength of the hidden-photon field (this applies to all cosmological scales of interest), it is appropriate to replace the above expression with its expectation value in that region. 
We can then Fourier transform back into $k$-space, and write the energy density as
\begin{align}
\rho(t) &= \int \! d \ln k \, \frac{1}{2 a^2} \bigg(
 \frac{a^2 m^2}{k^2 + a^2 m^2} \mathcal P_{\partial_t \A_L}(k, t) + m^2 \mathcal P_{\A_L}(k, t)   \bigg) \, ,
\label{eqn: energy-density}
\end{align}
where the power spectrum of a (homogeneously and isotropically distributed) field is defined by 
\begin{equation}
\langle X(\vec k, t)^* X(\vec k', t)\rangle \equiv (2 \pi)^3 \delta^3(\vec k-\vec k') \frac{2 \pi^2}{k^3} \mathcal P_X(k,t) \, ,
\label{eq:power-spectrum}
\end{equation}
so that $\langle X^2 \rangle = \int \! d \ln k \, \mathcal P_X(k,t) $ .

\subsection{Generation of inflationary fluctuations}
\label{sec:inflationary-fluctuations}

In the previous section, we saw that during inflation, until the modes were well outside the horizon, the action for the longitudinal modes is identical to that of a massless scalar field (see figure~\ref{fig:mode-evolution-1}), up to the simple rescaling Eq.~(\ref{eq:NGB-field-redef}). This means we can directly apply the standard results for a massless scalar to the longitudinal modes of the vector to understand the behavior of these modes during this phase. Light scalar fields are coherently produced during inflation.  Beginning their lives as small-scale vacuum fluctuations, Fourier modes grow in amplitude as they are expanded beyond the inflationary horizon, and thereafter are locked in as classical field fluctuations.\footnote{More technically, the growth of the scale factor $a(t)$ introduces an explicit time dependence into the field's action. The early-time vacuum therefore does not evolve into the late-time vacuum, but into a highly populated state, which can be found by performing a Bogolyubov transformation on the initial state.}. For a free, real, massless, canonical scalar field $\pi$, it is a standard result that modes exiting the horizon are described by
\begin{equation}
\pi(\vec k, t) = \pi_0(\vec k) \Big(1- \frac{i k}{a(t) H_I}\Big)e^{\frac{i k}{a(t) H_I}} \, ,
\label{eq:scalar-horizon-crossing}
\end{equation}
where $\pi_0(\vec k)$ is gaussian-distributed with power spectrum (as defined in Eq.~(\ref{eq:power-spectrum}))
\begin{equation}
\mathcal P_{\pi_0} (k) = \bigg(\frac{H_I}{2\pi}\bigg)^2 \, .
\label{eq:scalar-infl-fluc}
\end{equation}
Here $H_I$ is the value of the inflationary Hubble parameter (assumed to be slowly varying) around the time the modes exit the horizon. 

Using this result for the longitudinal modes of the vector and using the rescaling Eq.~(\ref{eq:NGB-field-redef}), we see that the amplitude in the longitudinal mode is 
\begin{gather}
\A_{L, \rm exit}(\vec k,t) = \A_0(\vec k) \Big(1- \frac{i k}{a(t) H_I}\Big)e^{\frac{i k}{a(t) H_I}}
\label{eq:long-horizon-crossing}
\\
\mathcal P_{\A_0} (k) = \bigg(\frac{k H_I}{2\pi m}\bigg)^2 \, .
\label{eq:long-infl-fluc}
\end{gather}

From Eq.~(\ref{eqn: energy-density}), we can see that the modes of a given $k$, at the point when they exit the horizon (i.e. when $k= a H_I$), contribute to the energy density as
\begin{equation}
\left. \frac{d \rho}{d \ln k} \right|_{\rm exit} \approx \frac{H_I^4}{(2\pi)^2}
\end{equation}

\subsection{Evolution after inflationary production}
\label{sec:evolution}
In this section, we compute the evolution of the fluctuations after they exit the horizon during inflation as well as their evolution from super-horizon modes to sub-horizon modes after the end of inflation. Once the modes exit the horizon during inflation, they evolve as  coherent classical field modes. The power spectrum of the field then evolves as
\begin{equation}
\mathcal P_{\A_L} (k, t) = \mathcal P_{\A_0} (k) \times \bigg(\frac{\A_L(\vec k, t)}{\A_0(\vec k)} \bigg)^2
=  \bigg(\frac{k H_I}{2\pi m}\bigg)^2 \times \bigg(\frac{\A_L(\vec k, t)}{\A_0(\vec k)} \bigg)^2 \, ,
\label{eq:power-spectrum-evolution}
\end{equation}
where $\A_L(\vec k, t)$ is the solution of the classical equations of motion with initial condition given by Eq.~(\ref{eq:long-horizon-crossing}).

\begin{figure}[h]
\begin{center}
\includegraphics[width=0.95\textwidth]{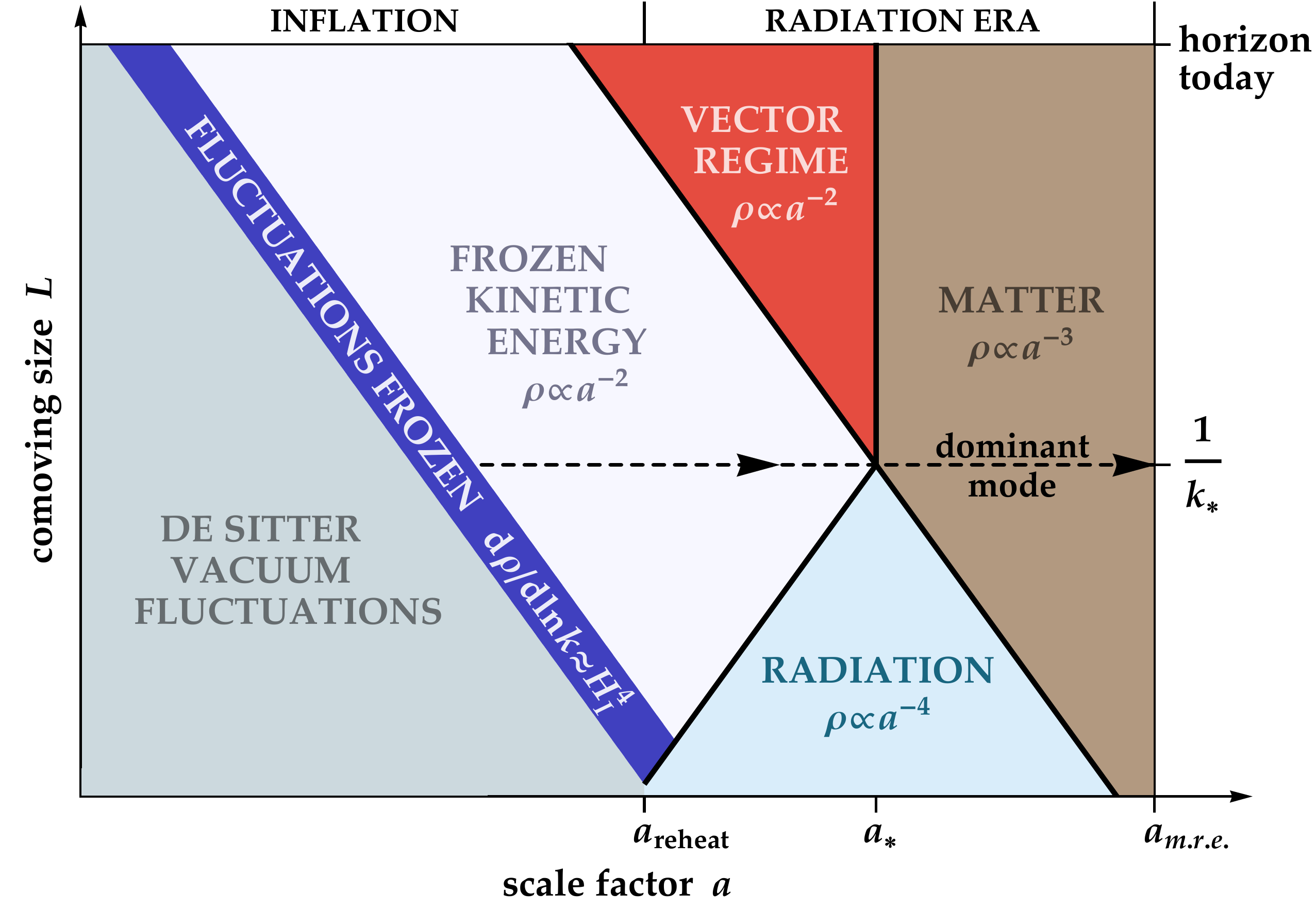}
\caption{Evolution of the energy density in longitudinal modes, from inflationary production through to matter radiation equality. As expected from figure~\ref{fig:mode-evolution-1}, the evolution is the same as it would be for a massive scalar in all regions except the red triangle labelled ``Vector regime''.  In that regime, the energy stored in a scalar would be constant, whereas for the vector it damps as $a^{-2}$.  This damping suppresses large-scale isocurvature modes, allowing the produced vector abundance to make up the dark matter.  This abundance is dominated by modes of comoving size $1/k_*$ indicated by the dashed line. 
The details of reheating do not affect these modes, as long at it occurs before they reenter the horizon.
}
\label{fig:mode-evolution-2}
\end{center}
\end{figure}

The equation of motion for the longitudinal modes follows trivially from their action:
\begin{gather}
\Big(\partial_t^2 + \frac{3 k^2 + a^2 m^2}{k^2 + a^2 m^2} H \partial_t + \frac{k^2}{a^2} + m^2 \Big) \A_L = 0 \, .
\label{eq:long-eom}
\end{gather}
We now solve this analytically in different regimes.  The first three regimes will correspond to the pale blue shaded region of figure~\ref{fig:mode-evolution-1}, and we will indeed see identical behavior to a free massless scalar. The fourth regime we consider corresponds to the brown shaded region of figure~\ref{fig:mode-evolution-1}, and there as expected we will see matter-like behavior.  Finally we turn to the red shaded, ``uniquely vector-like'' region of figure~\ref{fig:mode-evolution-1}, where we will discover new behavior unlike that of a scalar. 
We summarize the results in figure~\ref{fig:mode-evolution-2}.

\begin{itemize}

\item \textbf{Horizon exit during inflation} -- $H = H_I \gg m$. We can drop $m^2$, and the equation of motion is approximately
\begin{gather}
\big(\partial_t^2 + 3 H_I \partial_t  +k^2/a^2\big) \A_L \approx 0 
\quad \Longleftrightarrow \quad
( a \partial_a a \partial_a + 3 a\partial_a +k^2/(a H_I)^2) \A_L  \approx 0\, .
\label{eq:superhorizon-relativistic-eom}
\end{gather}
The solution to this a term of the form Eq.~(\ref{eq:long-horizon-crossing}), plus second term whose form is the complex conjugate of the first.  We see that the inflationary initial condition (i.e. de Sitter vacuum) gives rise to only the first term, setting the second to zero.

\item \textbf{Super-horizon relativistic regime} -- $H \gg k/a \gg m$. This is the region labelled ``frozen kinetic energy'' in figure~\ref{fig:mode-evolution-2}. Here we can drop the $k^2/a^2$ and $m^2$ terms, and the equation of motion is approximately
\begin{gather}
\big(\partial_t^2 + 3 H \partial_t \big) \A_L \approx 0 
\quad \Longleftrightarrow \quad
(\partial_a a H \partial_a + 3 H \partial_a ) \A_L  \approx 0\, .
\label{eq:superhorizon-relativistic-eom}
\end{gather}
During inflation, when \,$H = H_I \approx \text{const}$\,, this has solution
$\A_L = c_1 + c_2 a^{-3}$, 
while during radiation domination, when \,$H \propto a^{-2}$\,, this has solution
$\A_L = c_1 + c_2 a^{-1} $.
In either case the second term dies fast, and so in this regime
\begin{equation}
\A_L = \A_0 = \text{const}
\end{equation}
to an extremely high accuracy.  
The contribution to the energy density from these modes therefore evolves as
\begin{equation}
\rho \sim m^2 \A_L^2/a^2 \propto a^{-2} \, .
\end{equation}
This behaviour is the same as it would be for a massless scalar, whose kinetic energy damps as $a^{-2}$  in this regime while the field itself is frozen. This is just as expected from the discussion in section~\ref{sec:vector-in-FRW}.

\item \textbf{Sub-horizon relativistic regime} -- $k/a \gg m, H$. This is the region labelled ``radiation'' in figure~\ref{fig:mode-evolution-2}. Here we can drop the $m^2$ terms, and treat the $H \partial_t$ term as a small perturbation. Switching to conformal time $d t = a d \eta$, the equation of motion is approximately
\begin{gather}
\big(\partial_t^2 + 3 H \partial_t +k^2/a^2 \big) \A_L \approx 0 
\quad \Longleftrightarrow \quad
(\partial_\eta^2 + k^2 + 2 a H \partial_\eta) \A_L  \approx 0\, ,
\label{eq:subhorizon-relativistic-eom}
\end{gather}
with solution
\begin{equation}
\A_L \approx \frac{1}{a} \big( c_1 e^{i k \eta} + c_2 e^{-i k \eta} \big) \, .
\end{equation}
The energy density in these modes evolving as radiation,
\begin{equation}
\rho \sim m^2 \A_L^2/a^2 \propto a^{-4} \, ,
\end{equation}
which again is the same as the well known behaviour of a massless scalar in this regime.

\item \textbf{Late time non-relativistic regime} -- $m \gg k/a, H$. This is the region labelled ``matter'' in figure~\ref{fig:mode-evolution-2}. Here we can drop the $k/a$ terms, and treat the $H \partial_t$ term as a small perturbation. The equation of motion is approximately
\begin{gather}
\big(\partial_t^2 + H \partial_t + m^2 \big) \A_L \approx 0 \, ,
\label{eq:late-nonrelativistic-eom}
\end{gather}
with solution
\begin{equation}
\A_L \approx \frac{1}{\sqrt a} \big( c_1 e^{i m t} + c_2 e^{-i m t} \big) \, .
\end{equation}
The energy density in these modes evolves as matter
\begin{equation}
\rho \sim m^2 \A_L^2/a^2 \propto a^{-3} \, ,
\end{equation}
just as it would for a massive scalar in this regime.

\item \textbf{Hubble-damped, non-relativistic regime} -- $H \gg m \gg k/a$. This is the region labelled ``vector regime'' in figure~\ref{fig:mode-evolution-2}. We can drop the $k^2/a^2$ and $m^2$ terms, and the equation of motion is approximately
\begin{gather}
\big(\partial_t^2 + H \partial_t \big) \A_L \approx 0 
\quad \Longleftrightarrow \quad
(\partial_a a H \partial_a + H \partial_a ) \A_L  \approx 0\, .
\label{eq:damped-nonrelativistic-eom}
\end{gather}
During inflation, when \,$H = H_I \approx \text{const}$\,, this is solved by 
$\A_L = c_1 + c_2 a^{-1}$.
As in the super-horizon relativistic regime, we can drop the rapidly-decaying second term. 
However, during radiation domination, when \,$H \propto a^{-2}$\,, the solution is
\begin{equation}
\A_L = c_1 + c_2 a
\end{equation}
Now there is a growing term, which one might guess would dominate the solution. However, this regime follows a long period of the super-horizon relativistic regime, in which the field became constant to an extremely high degree.  The continuity of $\A$ and $\partial_a\A$ then prevents the linearly-growing term from taking off in the current regime, and the correct solution is again
\begin{equation}
\A_L = \A_0 =  \text{constant} \, .
\end{equation}
This result is not entirely obvious, and needs checking with care -- we do this in appendix~\ref{sec:vector-regime-careful}.
The evolution of the energy density in these modes damps as $a^{-2}$.  
This is quite different to the energy density of massive scalar modes in this regime.  In that case, the field $\phi$ would be frozen, contributing to a constant vacuum energy $m^2 \phi^2$.
\begin{equation}
\rho \sim 
\begin{cases}
m^2 \A_L^2/a^2 \propto a^{-2}  \quad&\text{vector}
\\ m^2 \phi^2 = \text{const}  \quad&\text{scalar}
\end{cases}
\label{eq:scalar-vector-difference}
\end{equation}

\end{itemize}

We can now see that the power spectrum of the longitudinal modes of vectors produced by inflation is different from those of scalars. For both the vector and the scalar, modes of all values of $k$ have the same energy density when they exit the horizon, $d \rho / d \ln k \sim H_I^4$.  The subsequent redshift of this energy density varies for different modes and at different times as indicated in figure~\ref{fig:mode-evolution-2}. In particular, the energy density in vectors and scalars redshifts identically for modes that are in the pale blue and brown regions of figure~\ref{fig:mode-evolution-2}. This is unsurprising since in these regions, the effective action of the vector is identical to that of scalars (with some field redefinitions). However, in the red-region of figure~\ref{fig:mode-evolution-2}, as seen in \eqref{eq:scalar-vector-difference}, the energy density in vectors redshifts as $a^{-2}$ while the energy density in a scalar would be constant. 

For the vector, it can be seen fairly simply (geometrically) from figure~\ref{fig:mode-evolution-2} that the modes whose energy redshifts the least are those of the special wavenumber $k_*$. Modes with longer wavelengths (further up the figure) receive extra redshifting in the ``vector regime'', causing the final energy in them to fall as $k^2$ at small $k$. Modes with shorter wavelength (further down the figure) undergo rapid redshifting while they are behaving like radiation, causing the final energy density in them to fall as $k^{-1}$ at large $k$. This means that the power spectrum of the vector has a peaked structure (shown in figures \ref{fig:field-power-spectra} and \ref{fig:density-power-spectra}), with power concentrated at the wavenumber $k_*$. This implies that the power in experimentally probed (CMB) long wavelength fluctuations  of the energy density of vectors is suppressed. This production mechanism does imply significant power at short wavelengths, but these have not yet been accessed by experiment. 

This is in stark contrast to scalars. The energy density in a scalar mode is constant when it is in the red-region of figure~\ref{fig:mode-evolution-2}. Hence,  all modes with momentum lower than $k_*$ have the same power, leading to isocurvature power at length scales probed by the CMB.

\subsection{Final relic abundance}
\label{sec:final-abundance}

\begin{figure}[t]
\begin{center}
\includegraphics[width=0.75\textwidth]{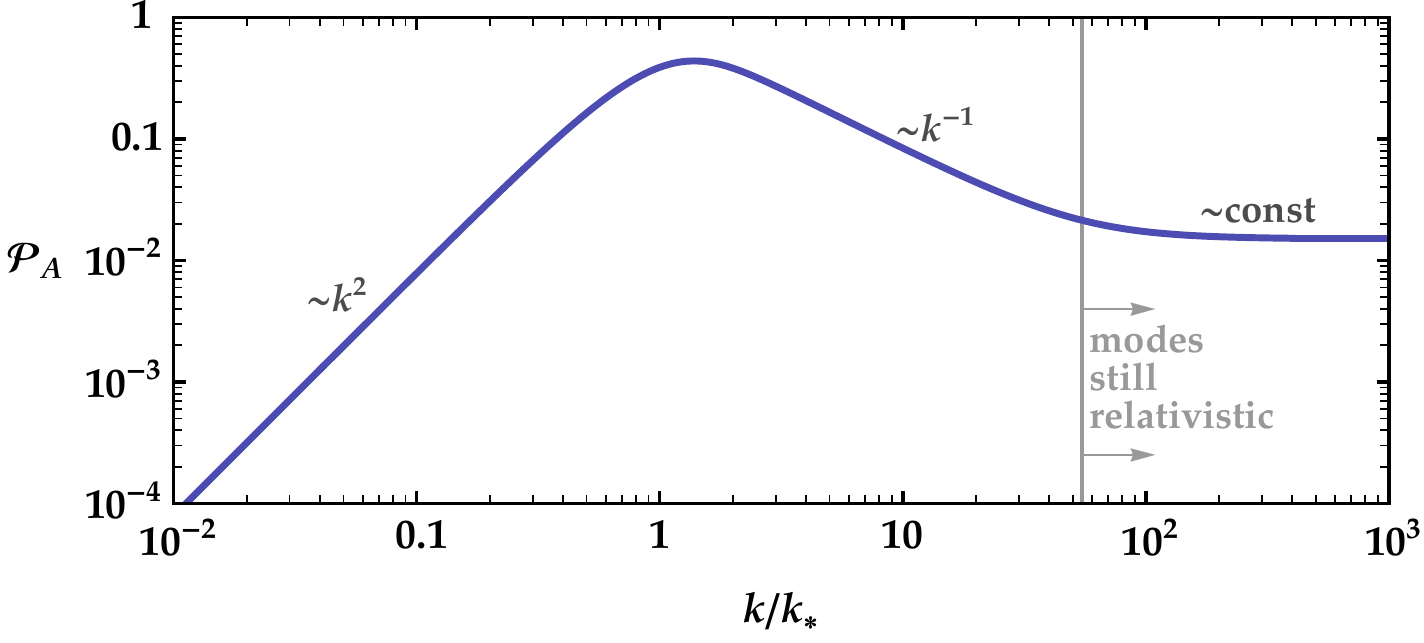}
\caption{Primordial power spectrum of the amplitude of a massive vector field's longitudinal modes, produced purely by inflationary fluctuations. 
The spectrum is shown at a time just 3 e-folds after $H=m$, and shorter wavelength modes (on the right of the plot) are still relativistic.
At later times the $k^{-1}$ scaling will continue all the way to the right of the plot.
}
\label{fig:field-power-spectra}
\end{center}
\end{figure}

We can now make use of the scalings described in the previous subsection, and calculate the final abundance of cold vector matter that was produced.  As shown in figure \ref{fig:field-power-spectra}, the power spectrum of the massive vector is peaked at the wavenumber $k_*$. This means we can estimate the final abundance by considering only the contribution from $k\approx k_*$. The energy density starts as $H_I^4/(2\pi)^2$ at horizon exit, redshifts as $a^{-2}$ until $a=a_*$, and then redshifts like matter. $a_*$ is the value of $a$ when $H=m$, and $k_* = a_* m$. These are approximately related to Hubble at matter-radiation equality by
\begin{equation}
a_* = a|_{H=m}\approx \sqrt \frac{H_{\rm m.r.e.}}{m} \times a_{\rm m.r.e.} 
\qquad \qquad
k_* = a_* m \approx \sqrt {H_{\rm m.r.e.} m} \times a_{\rm m.r.e.} \, .
\label{eq:astar-and-kstar}
\end{equation}
This gives an estimate of the abundance at matter-radiation equality
\begin{gather}
\rho_{\rm vector} 
\approx \frac{H_I^4}{(2 \pi)^2} \Big(\frac{a_{\rm exit}}{a_*}\Big)^2  \Big(\frac{a_*}{a}\Big)^3
\approx \frac{H_I^2 H_{\rm m.r.e.}^{\frac{3}{2}} m^{\frac{1}{2}} }{(2 \pi)^2} \Big(\frac{a_{\rm m.r.e.}}{a}\Big)^3 \, ,
\end{gather}
or, using $H_{\rm m.r.e.}^2 \approx \rho_{\rm cdm}(a_{\rm m.r.e.}) / M_{Pl}^2 \approx T_{\rm m.r.e.}^4/ M_{Pl}^2$,
\begin{gather}
\frac{\Omega_{\rm vector}}{\Omega_{\rm cdm}}
\approx \frac{H_I^2 m^{\frac{1}{2}} }{(2 \pi)^2 M_{Pl}^\frac{3}{2} T_{\rm m.r.e.}}\, .
\end{gather}

To get a more precise result, we can combine Eqs.~(\ref{eqn: energy-density}, \ref{eq:power-spectrum-evolution}) to write the vector abundance as
\begin{align}
\rho_{\rm vector}(t) &=  \frac{m^2}{2a^2}   \int \! d \ln k \, \Big( \frac{a^2}{k^2+a^2m^2}\mathcal P_{\partial_t \A_L}(k, t) + \mathcal P_{\A_L}(k, t) \Big)
\\
&=  \frac{H_I^2}{8 \pi^2} \frac{a_* k_*^2}{a^3}  \times  \int \! d \ln k \, \frac{k^2}{k_*^2} \,
\bigg( \frac{a}{a_*}\bigg| \frac{\A_L(k, t)}{\A_0(k)}\bigg|^2
+ \frac{a^3}{a_*(k^2+a^2m^2)} \bigg|\frac{\partial_t \A_L(k, t)}{\A_0(k)}\bigg|^2 \bigg)\, .
\end{align}
Here $\A_L(k, t)$ is the solution to the classical equations of motion, with initial conditions at horizon exit given by Eq.~(\ref{eq:long-horizon-crossing}). 
The integral is an $\mathcal O(1)$ factor which can be found by numerically solving for $\A_L(k, t)$. At late times it is independent of $m$, and we find it to be approximately equal to $1.2$. 
The $\sqrt m$ scaling comes from the $a_* k_*^2$ factor, which can be calculated precisely from the definitions above. 
Given the observed dark matter abundance, $\rho_{\rm cdm}(t_0) = 1.26 \times 10^{-6} \, \text{GeV}/\text{cm}^3$, we find the final result
\begin{equation}
\frac{\Omega_{\rm vector}}{\Omega_{\rm cdm}} = \sqrt \frac{m}{6 \times 10^{-6} \, \text{eV}} \,  \bigg(\frac{H_I}{10^{14} \, \text{GeV}} \bigg)^2 \, .
\label{eq:final-abundance}
\end{equation}

\subsection{Comment on misalignment production}

Previous work has considered production of a vector dark-matter abundance from cosmological initial conditions~\cite{Nelson:2011sf, Arias:2012az, An:2014twa}.  Our results are different from the results of this previous work.  The reasons for the differences are as follows.  References \cite{Arias:2012az, An:2014twa} added a specific coupling to the scalar curvature in the lagrangian, while we do not include this term.  Reference~\cite{Arias:2012az} showed that, in contrast to the case of a scalar, misalignment is ineffective at producing a vector abundance, unless this large coupling to the scalar curvature is added.  We agree with this result.  Further we note that in this case, inflationary fluctuations cannot produce the dark matter, so one must use misalignment.  There is in general both misalignment production and production by inflationary fluctuations.  When the coupling to the scalar curvature is added so that the massive vector behaves as a scalar, inflationary fluctuations will produce the normal, flat spectrum, just as for a scalar.  However if this makes up all of the dark matter it is ruled out by the bound on isocurvature perturbations by many orders of magnitude.  Thus in the case where the coupling to the scalar curvature is added, misalignment production (or some other mechanism) must dominate over production from the inflationary fluctuations by orders of magnitude.  Additionally, we note that the large coupling to the curvature is not consistent with the treatment of $m^2$ as a small spurion of gauge symmetry breaking, and would be expected to introduce a quadratic divergence to the mass.

As noted in \cite{Arias:2012az}, without the large coupling to the scalar curvature, misalignment production is ineffective and cannot generate the dark matter abundance.  
This also applies to any long wavelength mode produced by inflationary fluctuations, as can be seen from the low $k$ part of Figure \ref{fig:field-power-spectra}.
We suspect however that it may be possible to have something like misalignment production of a massive vector, in a different way.
The norm of the vector decreases during inflation as we have shown, $g^{\mu \nu} A_\mu A_\nu \propto a^{-2}$.  Thus if we want a uniform (misalignment produced) massive vector field to be dark matter today, then at some point during inflation this norm would have passed very large scales (e.g.~the Planck scale), and the energy density also would have been large (larger than the total inflationary energy density).  This is another way to state why misalignment production does not work in the standard case.
However it is reasonable to suspect that the effective field theory of this massive vector breaks down at some point.  So beyond some field excursion the potential may change from a simple quadratic term.  In fact it may not be described as a simple, 4D field theory at all -- for example it may arise from a higher dimensional string theory.  Such a degree of freedom may naturally redshift very differently than a massive vector during inflation and we would need the full UV completion.  Clearly then the initial condition of when during inflation the massive vector starts from its effective cutoff is unknown but would determine the final abundance and may allow misalignment production.  This is a change to the massive vector at some point in its field space.  There may also be a dynamical change to it in the early universe either during inflation or radiation dominance.  If the massive vector is produced from some other field (e.g.~the way axions can be produced from strings) then this may create a dark matter abundance of the vector in a very different way which is not subject to these objections to misalignment production.  
Thus, while misalignment production does not work for the canonical massive vector field considered here, we note for completeness that there may well be other production mechanisms which make other predictions.

\section{Adiabatic and isocurvature density fluctuations}
\label{sec:fluctuations}

In order to be a good dark matter candidate, the massive vector must have the correct, observed power spectrum of the matter density of our universe.  In particular, it must not have large isocurvature perturbations on long length scales.  This is normally an issue for a light, bosonic field, but as we discuss in Section \ref{Sec: isocurvature} the massive vector naturally avoids this problem.  Further, it must have the observed adiabatic, nearly scale-invariant fluctuations on cosmological length scales.  We demonstrate in Section \ref{sec: adiabatic} that this is indeed the case.

\begin{figure}[t]
\begin{center}
\includegraphics[width=0.75\textwidth]{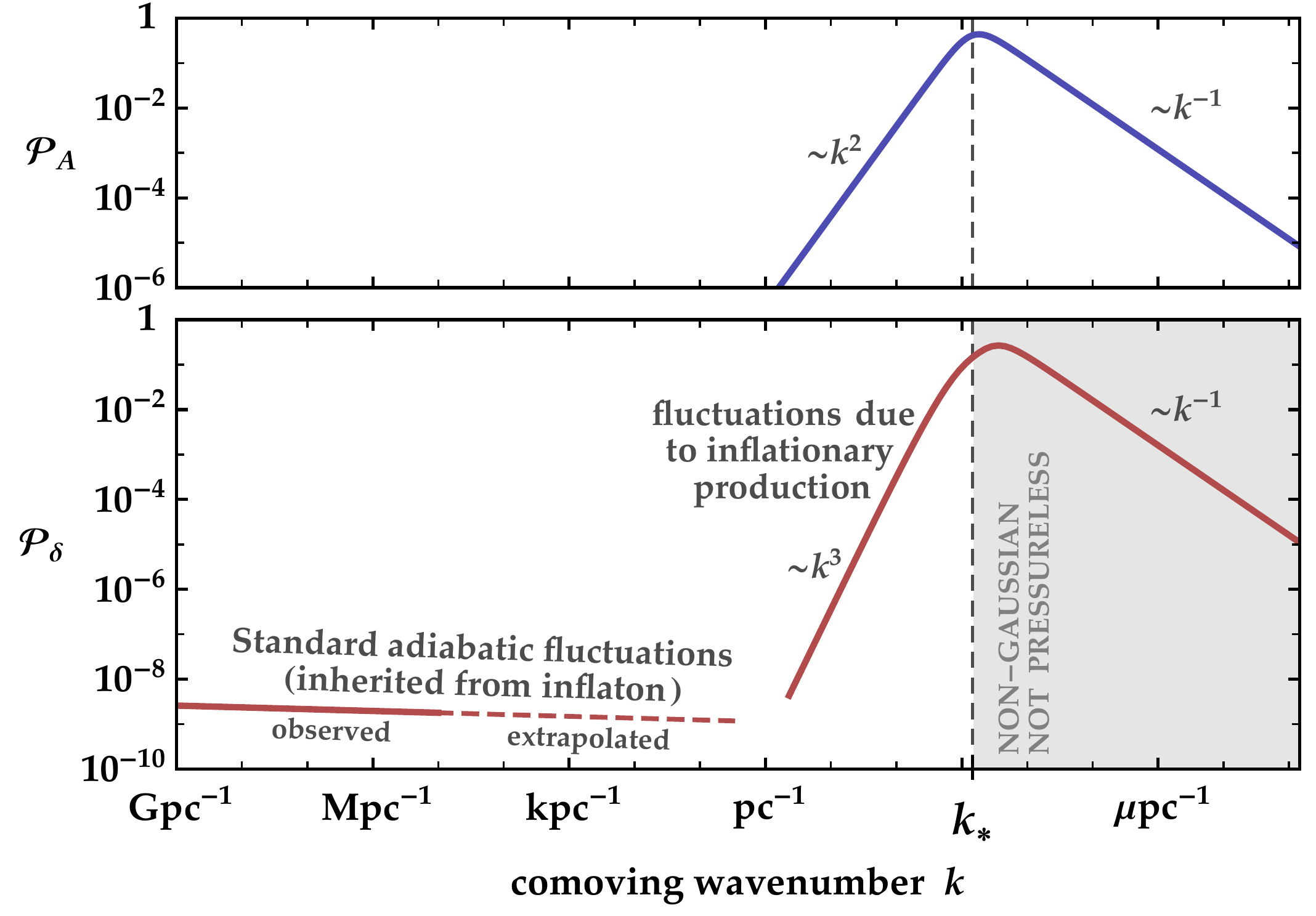}
\caption{(Lower plot) Primordial density power spectrum of a massive vector produced by inflationary fluctuations.
The spectrum is shown at a time when all modes covered are non-relativistic, but before self-gravitation of the modes is important (this should be approximately valid until matter-radiation equality). 
The peaked part of the spectrum at large $k$ is the isocurvature power produced by the inflationary fluctuations of the field itself. The also flat part of the spectrum at small $k$ corresponds to the usual adiabatic fluctuations, which are imprinted onto the field from inflaton fluctuations.
The values $m=10^{-5}\,$eV and $H_I \approx 10^{14}\,$GeV were used here, corresponding to $k_* \approx 1400\,\text{pc}^{-1}$. 
Because the density fluctuations are $\mathcal O(1)$ for $k\sim k_*$, fluctuations on shorter scales (in the gray region) are not well described by the power spectrum alone (see Fig.~\ref{fig:field-profile}).
These higher $k$ modes are also expected to be affected by quantum pressure later in their evolution. 
For comparison, the top plot shows the power spectrum of the field amplitude (figure~\ref{fig:field-power-spectra}).
}
\label{fig:density-power-spectra}
\end{center}
\end{figure}

\subsection{Isocurvature Fluctuations}
\label{Sec: isocurvature}

As demonstrated in Section \ref{sec:relic-abundance} the power spectrum of the massive vector field is peaked as in figure \ref{fig:field-power-spectra}, with $\mathcal O(1)$ power at the special scale $k_*$. 
Since they are generation without correlated fluctuations in the radiation bath, these are isocurvature fluctuations.
Isocurvature perturbations are dangerous, since on cosmological scales are strongly constrained by CMB observations.
However, this limit only applies to scales within a few orders of magnitude of the current size of the universe. 
On the other hand, $k_*$ corresponds to a cosmologically tiny scale,
\begin{equation}
1/k_* \sim 10^{10}\,\text{km} \times \sqrt{\frac{10^{-5}\,\text{eV}}{m}} \, .
\end{equation}
The isocurvature power spectrum fall off on long wavelengths (low $k$), to unobservable levels. 
To demonstrate explicitly this we calculate the density power spectrum in appendix \ref{sec: app power spectrum}.  The result is plotted as the peaked spectrum in the lower panel of Figure \ref{fig:density-power-spectra}. 
The power falls as $k^3$ below $k_*$, becoming completely negligible on cosmological scales.
We conclude that isocurvature perturbations are not a problem for the massive vector produced by inflation.

\begin{figure}[t]
\begin{center}
\includegraphics[width=0.95\textwidth]{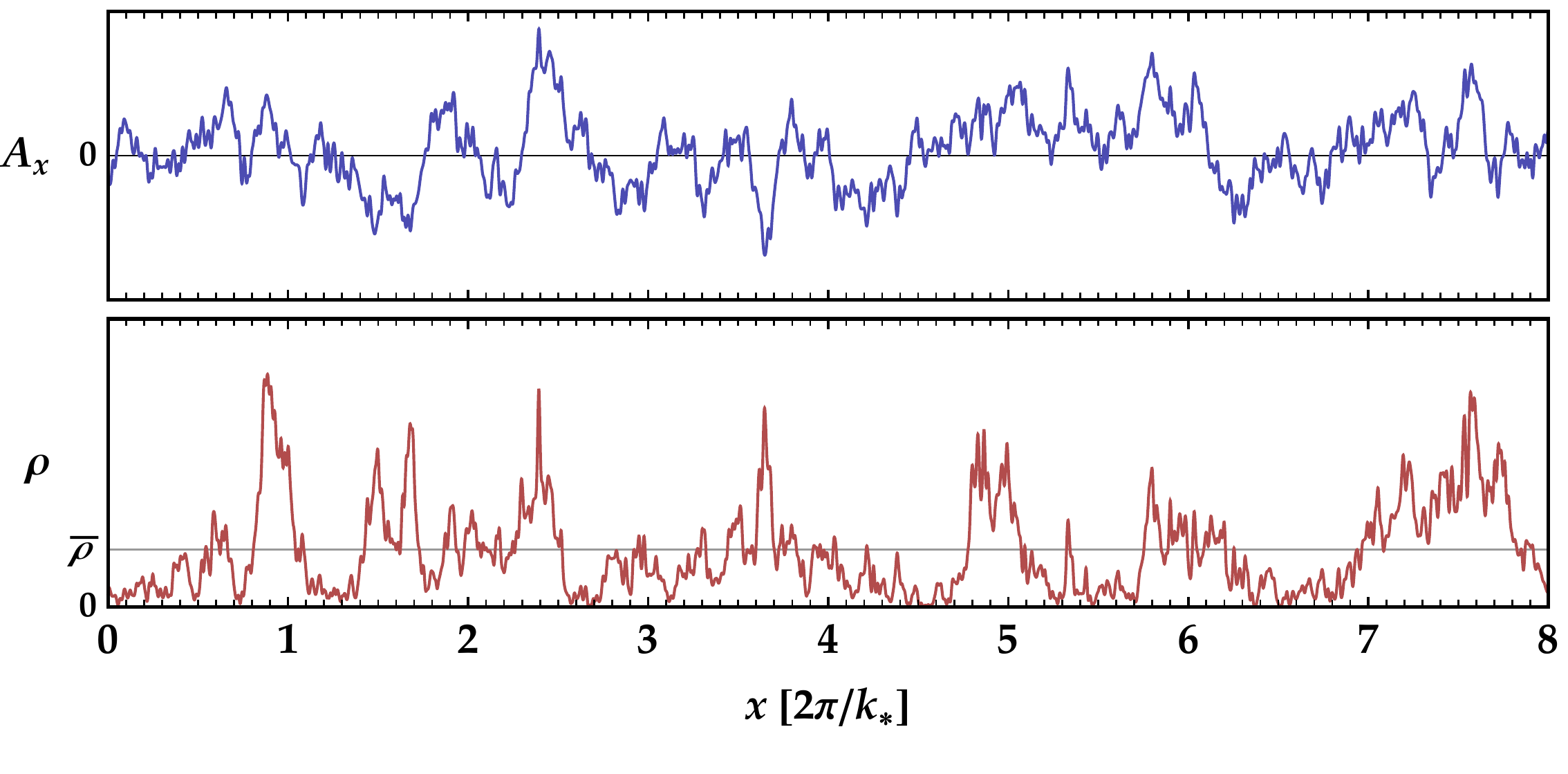}
\caption{Typical variation of the field and energy density along a line. This is a random example, generated according to the power spectra shown in Fig.~\ref{fig:density-power-spectra}, i.e. before structure formation occurs. 
(Note the $y$- and $z$-components of the field are not shown, and so the lower plot is not exactly the square of the top plot.)
As can be seen, the density is dominated by ``lumps'' of size $L\sim \pi/k_*$. 
On longer scales, overdensities are caused by random clustering of these lumps. 
On smaller scales, there is complicated sub-structure both within and between the main lumps. 
If the vector is discovered in direct detection experiments, its present-day field profile will be mapped out as the experiments sweep though the dark matter halo.
}
\label{fig:field-profile}
\end{center}
\end{figure}

To get a better picture of what this field looks like, one random realization of short-wavelength fluctuations of the vector field with the corresponding energy density is shown in Figure \ref{fig:field-profile}.  The $\mathcal O(1)$ fluctuations of the field and energy density are clear.  This could have interesting consequences for formation of dark matter structures on these length scales, as well as for direct detection experiments.  This is discussed further in Section \ref{Sec:Pheno}.

\subsection{Adiabatic Fluctuations}
\label{sec: adiabatic}

Intuitively the massive vector will pick up the adiabatic fluctuations of the inflaton because all of the abundance is generated by small wavelength fluctuations (small compared to cosmological scales).  Thus on much longer scales it looks homogeneous.  This is just like a WIMP, QCD axion, or any other dark matter candidate.  The distribution of WIMPs appears homogeneous on long scales but in fact there is large inhomogeneity at scales around the average inter-WIMP spacing.  Of course this does not affect the fact that the WIMP picks up the adiabatic fluctuations of the inflaton.  Although the peak in the massive vector power spectrum (found in Section \ref{sec:relic-abundance}) is at much longer scales than typical inter-WIMP spacings, it is still very short compared to cosmological scales.  So we know that the massive vector will indeed pick up the adiabatic fluctuations of the inflaton and so can be a good dark matter candidate.  In this subsection we show that this intuitive argument is correct.

We wish to calculate the spectrum of density perturbations imprinted on the massive vector field by the fluctuations of the inflaton.  The relevant inflaton fluctuations that are observed in the CMB are much longer wavelength than the modes of the massive vector that carry the dominant power.  This large separation of scales allows us to use the ``separate universes" approximation.  We will calculate the energy density of the massive vector assuming a homogeneous inflaton perturbation to the metric across the entire universe.  We follow the conventions of Dodelson~\cite{Dodelson:2003ft} and work in Newtonian Gauge and, as usual, ignore anisotropic stress. The perturbations are then described by substituting $g^{0 0} \to g^{0 0} (1 + 2\Phi)$ and $a^2 \to (1 + 2 \Phi)a^2$ in the unperturbed metric.  The perturbations  $\Phi(k, t)$ are zero during inflation, then turn on around reheating, and remain constant until they re-enter the horizon in the late universe.  Since $\Phi$ varies over wavelengths given by the size of the fluctuations seen in the CMB, and since these wavelengths are so much longer than the massive vector modes, $\Phi$ can be taken to be homogeneous.  Then we compare the energy density in different ``universes" (actually just different large regions of the universe) with different values of $\Phi$.   In this sign convention adiabatic perturbations to the matter density scale as $\rho \propto (1+ \frac{3}{2} \Phi )$.  By deriving this answer for the massive vector we will show that the fluctuations of the massive vector on the long length scales observed in the CMB are indeed adiabatic, with the correct, nearly scale-invariant spectrum arising from inflation. This allows the massive vector to be a good dark matter candidate.

Consider a region (or ``separate universe") with a metric perturbation.  We can find the action for the massive vector in this background by using the standard action for the longitudinal modes Eqn.~\eqref{eq:action-long}, reinstating $g^{0 0}$ with the perturbed value $-(1 + 2\Phi)$, and making the replacement $a^2 \to (1 + 2 \Phi)a^2$.  Thus, working only to linear order in $\Phi$, we find
\begin{gather}
S_{\rm Long} = \int \frac{a^3 (1 + 3 \Phi) d^3 k \, d t}{(2 \pi)^3 \sqrt{|g^{0 0}|} (1 + \Phi)} \,  \frac{1 - 2 \Phi}{2 a^2} 
\Big( \frac{(1 + 4 \Phi) g^{0 0} a^2 m^2}{k^2 + (1 + 2 \Phi) a^2 m^2} \big| \partial_t \A_L \big|^2 - m^2 \big| \A_L \big|^2   \Big) \, .
\label{eq:action-long2}
\end{gather}
This can be simplified by defining new variables
\begin{gather}
k' = k (1 - 2 \Phi)  
\label{eqn: k transform}
\\ 
m' = m (1 - \Phi).
\label{eqn: m transform}
\end{gather}
Then the action becomes
\begin{gather}
S_{\rm Long} = \int \frac{a^3 (1 + 3 \Phi) d^3 k \, d t}{(2 \pi)^3 \sqrt{|g^{0 0}|} (1 + \Phi)} \,  \frac{1}{2 a^2} 
\Big( - \frac{ a^2 {m'}^2}{{k'}^2 +  a^2 {m'}^2} \big| \partial_t \A_L \big|^2 - {m'}^2 \big| \A_L \big|^2   \Big) \, .
\label{eq:action-long3}
\end{gather}
Note the similarity with the usual action for the unperturbed massive vector Eqn.~\eqref{eq:action-long} (with $\Phi = 0$).  In fact, the equations of motion are exactly the same as for the unperturbed massive vector except with $k$ and $m$ replaced by $k'$ and $m'$, since the overall constant out front is not relevant for this.

Thus we  know  how the massive vector will evolve in this ``universe."  It will create a peaked power spectrum as shown above.  However the wavelength of the peak will change slightly, as will $a_*$, the scale factor at which the massive vector started acting like matter (as in Fig.~\ref{fig:mode-evolution-2}).  We now determine $a'_*$ by the scale factor when $H = m'$.  Since $H \propto a^{-2}$ during radiation dominance, we can relate this to the scale factor of matter-radiation equality $\amre$
\begin{gather}
\left( \frac{a'_*}{\amre} \right)^2 = \frac{\Hmre}{m'} = (1+\Phi) \frac{\Hmre}{m} 
\label{eqn: redshift factor for adiabatic}
\end{gather}
where $\Hmre$ is the Hubble constant at matter-radiation equality.  Note that matter-radiation equality is still defined globally to be the same time for all the ``separate universes" (so we do not put a prime on $\amre$).  This is because we want to know the energy density at a single globally defined time across the entire universe.  We are only concerned with the $\Phi$ dependence of our answers.  And the peak of the power spectrum for the massive vector occurs at $k' = k'_*$ which is defined by the crossing-point in Figure \ref{fig:mode-evolution-1} so we have $\frac{a'_*}{k'_*} = \frac{1}{m'}$.  This gives us a peak at
\begin{gather}
k'_* = \amre \sqrt{\Hmre m'} =  (1 - \frac{1}{2} \Phi) \amre \sqrt{\Hmre m}.
\end{gather}
Of course we want the actual wavelength at which the peak occurs, in other words we want the value of $k$ not the value of $k'$ at which the peak occurs.  These are related by Eqn.~\eqref{eqn: k transform}, so the peak is at
\begin{gather}
k = k_* = (1 + 2 \Phi) k'_* =  (1 + \frac{3}{2} \Phi) \amre \sqrt{\Hmre m}.
\label{eqn: modified k star peak}
\end{gather}
These formulae determine the power spectrum of the vector in this ``separate universe" with nonzero $\Phi$.

We also need to know the amplitude of the massive vector field produced during inflation in this universe.  The power produced during inflation is given by Eqn.~\eqref{eq:long-infl-fluc}, but now we need to modify it to take into account the nonzero $\Phi$.  We will only be interested in the power at the peak so we will use $k$ given in Eqn.~\eqref{eqn: modified k star peak}.  This does change with $\Phi$ because the wavelength of the peak in this ``universe" is physically different.  However in Eqn.~\eqref{eq:long-infl-fluc} we must use the unperturbed mass $m$ instead of $m'$.  This is because, in Newtonian Gauge, $\Phi$ is zero up until reheating where it the jumps to its constant value for the rest of the history of the universe.  Thus $\Phi = 0$ during inflation and so the power produced around the peak wavelength is
\begin{gather}
\mathcal P_{\A_0} (k_*) = \left(\frac{k_* H_I}{2\pi m}\right)^2 = (1 + 3 \Phi) \frac{\amre^2 \Hmre H^2_I}{(2 \pi)^2 m}
\label{eq:long-infl-fluc-adiabatic}
\end{gather}
After inflation the value of the field $A$ in the longitudinal mode with $k = k_*$ is then fixed until the crossing point $a'_*$, as in Figure \ref{fig:mode-evolution-2}.  Note that one could worry that this statement is no longer true at reheating because $\Phi$ jumps rapidly there.  However one can show from the equations of motion that this rapid change in $\Phi$ will not change the value of the field.  Thus from Eqn.~\eqref{eq:long-infl-fluc-adiabatic} we have the power in the dominant field mode when it begins to act like matter.

We are interested in the total energy density as a function of $\Phi$.  This is derived in the same way as Eqn.~\eqref{eqn: energy-density} was derived for the unperturbed universe.  But note that we want the energy density as seen by a global observer, not an observer in each ``separate universe."  Thus we use the global definition of position (and wavevector) so we Fourier transform in $k$ not $k'$ to find the energy density.  In the action for the longitudinal modes, Eqn.~\eqref{eq:action-long3}, the first factor in the integral is just the usual integral over phase space (in global coordinates).  Fourier transforming that action back into position space then gives an energy density
\begin{align}
\rho(t) &= \int \! d \ln k \, \frac{1}{2 a^2} \bigg(
 \frac{a^2 m'^2}{k'^2 + a^2 m'^2} \mathcal P_{\partial_t \A_L}(k, t) + m'^2 \mathcal P_{\A_L}(k, t)   \bigg) \, ,
\label{eqn: energy-density perturbed}
\end{align}
which is the same as Eqn.~\eqref{eqn: energy-density} except the $k$'s and $m$'s inside the integral have been changed to $k'$ and $m'$ (though note that the argument of the power $\mathcal P$ is not changed).  In the regime where the vector acts as cold dark matter, the kinetic energy and the potential energy are the same on average by the virial theorem.  So we can just take twice the second term for the average energy density:
\begin{align}
\rho(t) &= \int \! d \ln k \, \frac{1}{a^2} \bigg( (1 - 2 \Phi) m^2 \mathcal P_{\A_L}(k, t)   \bigg)  .
\label{eqn: energy-density perturbed2}
\end{align}
Finally the power at matter-radiation equality can be found in terms of the power produced by inflation Eqn.~\eqref{eq:long-infl-fluc-adiabatic} by redshifting the power by a factor $\frac{a'_*}{\amre}$ since the field redshfits $\propto \sqrt{a^{-1}}$ once it is acting as matter (as in Fig.~\ref{fig:mode-evolution-2}).  We will focus only on the power in the dominant mode so the integral is removed.  Using Eqns.~\eqref{eqn: redshift factor for adiabatic} and \eqref{eq:long-infl-fluc-adiabatic} in Eqn.~\eqref{eqn: energy-density perturbed2} then gives the energy density at matter-radiation equality
\begin{align}
\rho_\text{mre} \approx \left(1 + \frac{3}{2} \Phi\right) \frac{H^2_I \Hmre^\frac{3}{2} m^\frac{1}{2} }{(2 \pi)^2}  .
\label{eqn: energy-density perturbed3}
\end{align}
This is the correct result for adiabatic perturbations, that the energy density scales as $\left(1 + \frac{3}{2} \Phi\right)$.  All of our sign conventions can be checked by following the above procedure for a scalar instead of a vector, in which case one finds the same factor $\left(1 + \frac{3}{2} \Phi\right)$.  Thus the massive vector dark matter does indeed pick up the nearly scale-invariant, adiabatic perturbations of the inflation.  This is illustrated in the lower panel of Figure \ref{fig:density-power-spectra}.  So the massive vector can be a good dark matter candidate.

\section{Dark Matter phenomenology}
\label{Sec:Pheno}

\begin{figure}[t]
\begin{center}
\includegraphics[width=0.95\textwidth]{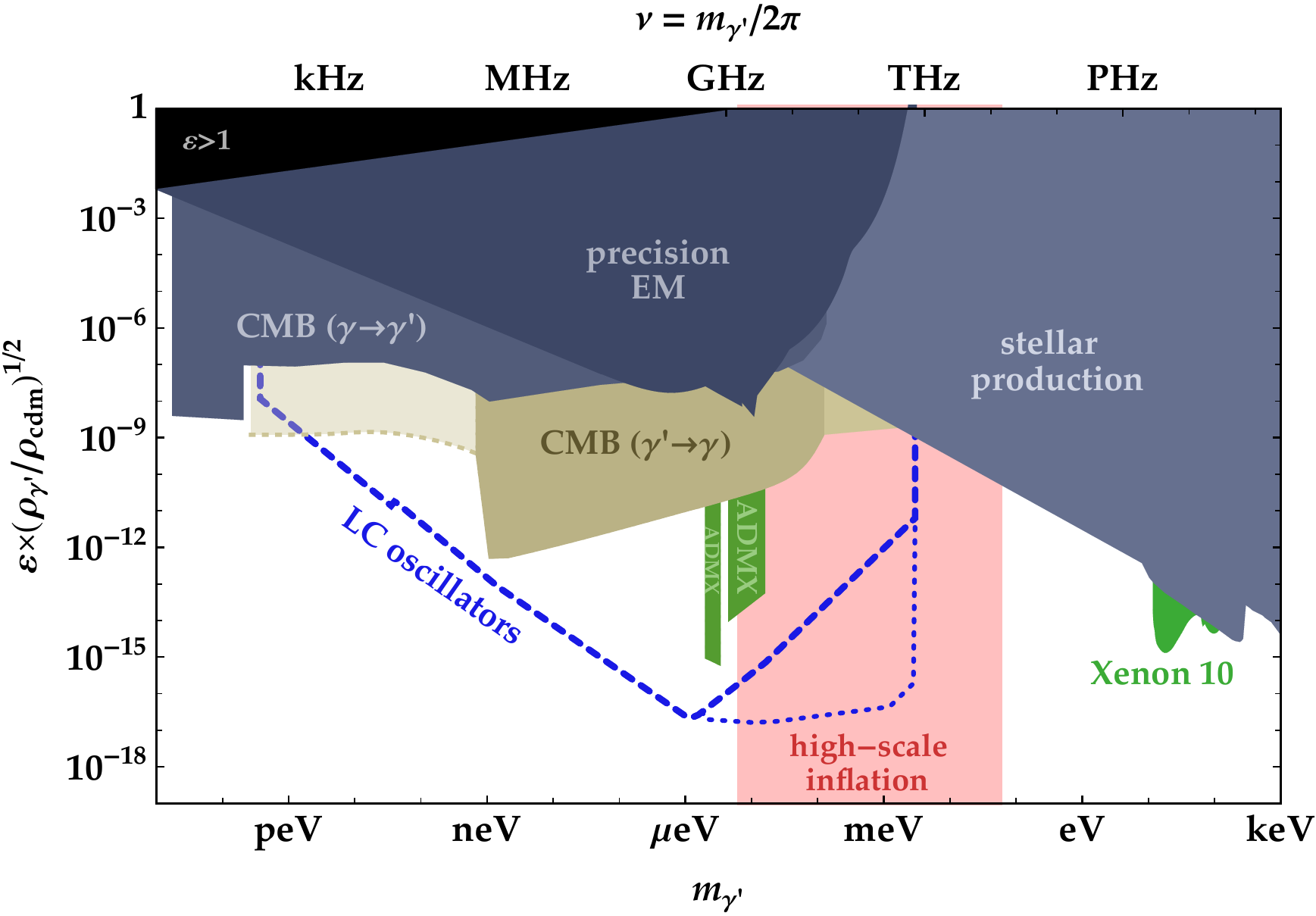}
\caption{Prospects for direct detection of a cosmological vector abundance through kinetic mixing with the photon. Direct detection experiments are sensitive to the combination $\varepsilon \sqrt{\rho_{\gamma'}}$ shown on the y-axis ($m_{\gamma'}$ is the mass of the new vector, and $\rho_{\gamma'}$ its cosmic abundance -- this plot does not assume the vector makes up all the dark matter).
In the vertical pink band, high-scale inflation ($10^{13}\,\text{GeV}\simlt H_I\simlt10^{14}\,\text{GeV}$) generates the full dark matter abundance for the vector. Masses to the right of this band require lower values of $H_I$, while to the left of this band the vector can constitute an interesting dark matter subcomponent.
The blue dashed line shows the projected reach of a recently proposed search with resonant LC oscillators~\cite{Chaudhuri:2014dla}, with the blue dotted line showing the potential improvement by multiplexing in the high-frequency regime. 
Tan shaded regions are excluded due to cosmological disturbances caused by the relic vectors, while green regions are excluded by null results from the Xenon10 and ADMX dark-matter searches. 
Gray-blue shaded regions indicate constraints on the existence of the vector independent of its cosmological abundance (these are truly constraints on $\varepsilon$, so to plot these here it was assumed that the vector makes up the largest possible fraction of the dark matter that is consistent with inflationary production).
See section~\ref{Sec:Pheno} for more details.}
\label{fig:direct-detection}
\end{center}
\end{figure}

We have shown that inflation automatically generates a cosmological abundance for any massive vector boson, with the abundance completely determined  by the Hubble scale during inflation and the mass of the vector boson. For masses $\sim 6\times 10^{-6} \text{ eV} \left(10^{14} \text{ GeV}/H_I\right)^4$, this abundance is equal to the observed dark matter density. For lower masses, it becomes a subdominant component of dark matter.  Given the model independent nature of this  mechanism, there is a strong case to search for this  abundance  through laboratory experiments. Detection in the laboratory requires the massive vector boson to interact non-gravitationally with the standard model.

It is reasonable to expect such interactions. For example, the massive vector boson could kinetically mix with the photon, though the operator
\begin{equation}
\mathcal L \supset \frac{1}{2} \varepsilon F_{\rm EM}^{\mu \nu} F'_{\rm \mu \nu} \, ,
\end{equation}
where here $F'_{\rm \mu \nu}$ represents the new vector's field strength. As a dimension four operator  allowed by symmetry, it is unsuppressed by heavy mass scales and is generically expected to exist, although different UV completions allow $\varepsilon$ to be naturally extremely small~\cite{Holdom:1985ag}. 
Such an interaction can lead to a wide variety of possible constraints and new searches -- see for example~\cite{Jaeckel:2010ni, Bjorken:2009mm, Ahlers:2007rd, Jaeckel:2007ch,  Wagner:2010mi, Povey:2010hs, Parker:2013fba, Betz:2013dza, Graham:2014sha, Pospelov:2007mp, Abel:2008ai, ArkaniHamed:2008qn, ArkaniHamed:2008qp, Pospelov:2008zw, Goodsell:2009xc, Arvanitaki:2009hb, Reece:2009un, Batell:2009di, Aubert:2009af, Essig:2010ye, deNiverville:2011it, Ringwald:2012hr, Hewett:2012ns, Dharmapalan:2012xp, Battaglieri:2014hga, Horns:2012jf, Dobrich:2014kda, Suzuki:2015sza, Redondo:2015iea} and other references given below in this section.
If so, over a wide range of masses $10^{-12} - 10^{-3}$ eV,  a cosmological abundance of such particles can be detected using well developed electromagnetic resonator technologies as suggested in~\cite{Arias:2012az, Chaudhuri:2014dla, Arias:2014ela}. The estimated reach of such experiments is plotted in figure \ref{fig:direct-detection}.  These experiments are sensitive to the combination $\varepsilon \sqrt{\rho_{\gamma'}}$ where $\rho_{\gamma'}$ is the local energy density in the massive vector boson, which of course, cannot be larger than the dark matter density. Since the sensitivity scales as $\sqrt{\rho_{\gamma'}}$, and since $\varepsilon$ is a priori unknown, these detectors have the ability to probe  vector boson densities that are a small subcomponent of the dark matter density. This is particularly important in this scenario since the abundance generated by inflation will be smaller than the dark matter density for vectors with mass less than  $\sim 10^{-5} \text{ eV} \left(10^{14} \text{ GeV}/H_I\right)^4$. Given our ignorance of the ultraviolet physics responsible for the origins of these massive vector bosons and the model independent way in which the vector boson abundance is generated, there is a strong case to search for such a cosmic abundance over a wide range of parameters. 

Interestingly, as is clear from figure  \ref{fig:direct-detection}, electromagnetic resonator technology is particularly well suited to search for massive vector bosons in a several-decade mass range around $\sim\!10^{-6}$ eV, corresponding to resonator frequencies $\sim 10^8$ Hz. If the scale of inflation is close to the present experimental bound of $H_I \sim  10^{14}$ GeV, a vector boson in this mass range will have an abundance equal to the observed dark matter density. There is thus strong motivation to leverage the existence of high precision electromagnetic sensors in this frequency range as a positive detection could yield strong evidence for inflation. 
The blue dashed line shows the projected reach of a recently proposed search with resonant LC oscillators~\cite{Chaudhuri:2014dla}, with the blue dotted line showing the potential improvement by multiplexing in the high-frequency regime. 

As is evident from figure \ref{fig:direct-detection}, there are a variety of constraints on such kinetically mixed vector bosons. A cosmic abundance of these vector bosons is constrained in the tan shaded regions. The central tan shaded region in figure \ref{fig:direct-detection} is ruled out by  CMB distortions due to conversion of the vector into photons~\cite{Arias:2012az}. The rightmost region by is excluded due to depletion of the vector by conversion into photons. We derive this bound following the analysis of~\cite{Arias:2012az}, but taking the initial abundance to be the maximum allowed by inflationary production, whereas in~\cite{Arias:2012az} an exponentially large initial abundance was considered. 
The leftmost, semi-transparent tan region is claimed in~\cite{Arias:2012az} to be excluded due to depletion of the vector by conversion into plasma modes. This analysis appears to have ignored the (potentially large) disruptive effect this would have on the baryon plasma, and so may not be valid. We do not attempt to reanalyze this constraint here.
As pointed out in~\cite{An:2014twa}, conventional dark matter direct detection experiments are sensitive to vector dark matter if its mass is above the experimental threshold energy. Similarly, as pointed out in~\cite{Arias:2012az}, the ADMX axion dark matter search is also sensitive to vector dark matter. The green shaded regions in~\ref{fig:direct-detection} show the corresponding exclusions from results from the ADMX~\cite{Asztalos:2011bm} and the Xenon10~\cite{Angle:2011th} experiments.

The blue-gray shaded regions are excluded regardless of the cosmic abundance of the vector boson -- they only rely on the existence of the vector boson as a degree of freedom in the theory.  These bounds include stellar production of the vector~\cite{An:2013yfc, An:2013yua}, precision tests of electromagnetism~\cite{Bartlett:1988yy, Betz:2013dza, Graham:2014sha}, and distortion of the CMB due to conversion of photons into the vector~\cite{Mirizzi:2009iz}. 
Since these are bounds on $\varepsilon$ rather than $\varepsilon \sqrt{\rho_{\gamma'}}$, an assumption about the vector abundance was needed in order to put the bounds on the plot.
In the vertical pink band and to its right, the full vector dark matter abundance can be generated by inflation consistently with bounds on the inflationary scale ($H_I \simlt 10^{14}\,$GeV), and we assume this to be the case. To the left of this band, we use the maximum abundance that can be generated by inflationary fluctuations ($\rho\propto \sqrt m$).  With this assumption, the black region in the upper left would require $\varepsilon > 1$ which is forbidden.

The primordial dark matter power spectrum produced by this mechanism (see figure \ref{fig:density-power-spectra})  is peaked  at comoving scales of order $k_*^{-1} \sim 10^{10}\,\text{km} \times \sqrt{10^{-5}\,\text{eV}/m}$, and drops as $k^3$ for larger comoving scales.
Thus, over a range of length scales near this peak, the density perturbations are significantly bigger than the adiabatic density fluctuations of $\sim 10^{-5}$. 
They can therefore be expected to go non-linear and become self bound before large-scale-structure (e.g. galaxy) formation, resulting in significant small-scale dark-matter substructure.\footnote{We estimate that field-gradient pressure may be significant for modes shorter than $k_*^{-1}$, possibly preventing their collapse, but that longer scale modes will behave as cold pressureless matter, allowing them to form structures.}
At the peak of the power spectrum, the density fluctuations are $\mathcal O (1)$, and hence we expect these fluctuations to become self-gravitating around matter-radiation equality. These peak fluctuations would then decouple from the subsequent expansion of the universe, leading to structures in the dark matter density today at distances $\sim$0.1\,AU, with over-densities of $\sim$$10^5 \, \text{GeV}/\text{cm}^{-3}$. 
Longer wavelength fluctuations, with lower power, can be expected to decouple somewhat later, leading to a range of 
dark-matter over-densities today on length scales not too far from 1\,AU -- i.e. roughly the distance the earth moves in a year. 
Structure on these  scales is therefore traversed by laboratory experiments over experimentally accessible time scales. 
Remarkably, direct detection experiments should therefore be able to directly probe this structure, measuring a range around the peak of the power spectrum. 
This would give dramatic evidence for this production mechanism, enabling a direct probe of inflationary dynamics through the physics of dark matter. For these ideas to bear fruition, it is necessary to understand the non-linear formation and effects of galactic dynamics on such structures. 
It would also be interesting to see if these structures have other observable astrophysical effects. 
 We will explore these consequences in future work.

\section{Conclusions}
\label{Sec:Conclusions}

We have shown that the observed dark matter abundance and the spectrum of its density inhomogeneities can naturally be produced by inflation if the dark matter particle was an ultra-light vector boson. 
The production mechanism is entirely gravitational -- it only requires the existence of a massive vector boson without significant self interactions. As long as the coupling between the standard model plasma and this vector boson is small, the abundance of the massive vector boson produced by inflation will not thermalize and will contribute to the dark matter density today.  Unlike scalars,  inflationary production of massive vector boson dark matter does not lead to large scale isocurvature perturbations on the CMB. Instead, the dark matter density has large isocurvature perturbations only at short distances that have not been probed by experiment. On the large distances probed by the CMB, it has suppressed isocurvature power and its power spectrum is dominated by the usual adiabatic density perturbations that are imprinted on it from inflaton fluctuations. 

The abundance produced by this mechanism is calculable -- it is completely determined by the Hubble scale during inflation (specifically, when the dominant modes exit the horizon) and the mass of the vector boson. 
This is in contrast to other production mechanisms that are often discussed for ultra-light bosonic dark matter candidates such as axions. For example, the density produced by the misalignment production mechanism \cite{Dine:1982ah, Preskill:1982cy} entirely depends upon unknown initial conditions. 
Moreover, the misalignment mechanism does not generate a significant cosmic abundance of massive vectors unless there are large  additional interactions. 
Unlike scalars, the energy density in the massive vector boson field redshifts rapidly during inflation, diluting the initial abundance in the field. While this dilution can be avoided through the introduction of additional finely tuned interactions~\cite{Arias:2012az}, or if the misalignment was generated near the end of inflation,  they add complexity to this scenario. 
Other  mechanisms such as the emission of these particles from topological defects such as strings may yield calculable abundances but are beset with computational challenges \cite{Hiramatsu:2012gg}. 
Further, it may be difficult to obtain direct observational evidence for such events in the early universe. If the Hubble scale during inflation is inferred through observations of the primordial gravitational wave spectrum, our mechanism predicts the approximate mass ($ m \approx 6\times 10^{-6} \,\text{eV} \left(10^{14} \,\text{GeV}/H_I \right)^4 $) a vector boson must have in order to constitute all of the dark matter of the universe.

If the inflationary Hubble scale was in the range $10^{13}$$-$$10^{14}$ GeV that is accessible to the next generation of primordial gravitational wave experiments, the predicted range of the vector boson dark matter mass would be $\sim$$10^{-5}$$-$$10^{-1}$\,eV. These vector bosons may interact with the standard model through kinetic mixing, enabling the possibility of direct detection. Of course,  in contrast to WIMP dark matter, since the inflationary production is gravitational, such interactions are not essential for this mechanism. 
However, if such interactions exist, massive vector boson dark matter can be searched for over this range of masses using well developed electromagnetic resonator technologies, as suggested in \cite{Chaudhuri:2014dla, Arias:2014ela}. Further, the existence of new vector bosons in this frequency range can also be directly probed by various ``Light-Shining-Through-a-Wall" experiments wherein these vector bosons are produced by a laboratory source and then subsequently detected \cite{Graham:2014sha, Dobrich:2013mja}. Thus, in this range of parameters, this scenario leads to an exciting set of experiments. If vector boson dark matter is detected in the mass range $10^{-5}$$-$$10^{-1}$ eV,  it would give strong impetus to the next generation of primordial gravitational wave detectors. 

This mechanism, being gravitational, will generate a cosmological abundance for a massive vector field, even when its mass is not $6\times 10^{-6} \,\text{eV} \left(10^{14} \,\text{GeV}/H_I \right)^4 $. For lower masses, the abundance generated will be subdominant to the dark matter density. It is important to devise experiments to search for such light vector boson dark matter even if they are a subdominant component of the dark matter density, since a discovery in such a channel provide a new experimental probe of inflation. 
Conversely, if a high inflationary scale is experimentally observed, our results would seem to constrain larger vector masses.
However, it should also be noted that our results are subject to assumptions about the behavior of the horizon around the time the dominant modes exit and reenter it. For example, the universe could have been reheated to a very low temperature following a period of matter domination. In that case, the dominant vector modes would begin to redshift like matter earlier, resulting in a lower final abundance, and allowing a significantly larger vector mass.

The  power spectrum of the dark matter density produced by this mechanism is peaked at comoving scales $k_*^{-1} \sim 10^{10}\,\text{km} \times \sqrt{10^{-5}\,\text{eV}/m}$. Dark matter perturbations at this comoving scale become self-gravitating  during the era of matter-radiation equality and decouple from the subsequent expansion of the universe, leading to structure in the dark matter density at distances $\sim 0.1$ AU.  It is thus possible to sample different parts of this structure over $\sim$year time-scales, 
and hence to directly measure it in the laboratory with direct-detection experiments. 
The detection of such short distance structure in the dark matter would give dramatic evidence for this production mechanism, and a would provide a new probe of inflation itself.

A central reason for the success of this production mechanism is the fact that the dark matter density inherits large distance adiabatic perturbations. Hence, even if the dark matter production mechanism has significant power at short distances, as long as the mechanism does not directly produce similar power at large distances, it will reproduce the observations of the CMB. It would be interesting to see if there are other classes of theories where such a suppression may naturally occur. These might enable the existence of structure in the dark matter at length scales ({\it e.g.} $\sim$1\,AU) that are short on galactic length scales but large enough to be relevant for laboratory experiments. 

Cosmic Inflation is theoretically compelling. Weakly interacting, ultra-light fields such as the graviton are naturally produced during inflation and as a consequence of their suppressed interactions, their abundance is not thermalized by the plasma. Gravitational waves are known to exist and are hence a natural target for experimental searches. However, in light of the challenges that must be overcome to detect them, it is interesting to ask if inflationary dynamics can be probed through the existence of other weakly interacting, ultra-light fields. Such fields exist naturally in many frameworks of physics beyond the standard model and they can interact with the standard model in a variety of ways, unlike the restricted interactions of gravitational waves. The remarkable advances in the field of precision metrology over the past two decades have made it possible for us to search for a cosmological abundance of these particles over a wide range of masses and interactions. A discovery in such an experiment would probe the ultra-high energy scales that are responsible for the production of these particles, the mechanics of inflation and the subsequent evolution of the universe.


\section*{Acknowledgements}
We would like to thank N.~Arkani-Hamed, L.~Dai, T.~Jacobson, M.~Kamionkowski, D.E.~Kaplan, and R.~Sundrum
for useful discussions. This work was supported in part by
NSF grant PHY-1316706, DOE Early Career Award DE-SC0012012, and the Terman Fellowship. SR acknowledges the support of NSF grant PHY-1417295.

\vspace{0.5 cm}

\appendix

\section{Power spectrum of the field and the density}
\label{sec: app power spectrum}

The power spectrum $\mathcal P_X (k, t)$ of a (homogeneously and isotropically distributed) field $X$ is defined here as
\begin{equation}
\langle X(\vec k, t)^* X(\vec k', t)\rangle = (2 \pi)^3 \delta^3(\vec k-\vec k') \frac{2 \pi^2}{k^3} \mathcal P_X(k,t) \, ,
\end{equation}
so that
\begin{equation}
\int \! d \ln k \, \mathcal P_X(k,t) = \langle X^2 \rangle  \, .
\end{equation}

At late times the energy density $\rho (\vec x)$ in the hidden photon field is proportional to $|\vec A(\vec x)|^2$. 
Defining the density fluctuation $\delta (\vec k)$ in the usual way,
\begin{equation}
\rho(\vec x) = \langle \rho \rangle \left( 1 + \int \frac{d^3 k}{(2 \pi)^3} \delta(\vec k) e^{i \vec k \cdot \vec x} \right) \, ,
\end{equation}
$\delta(\vec k)$ is given by
\begin{equation}
\delta (\vec k) = \frac{1}{\langle |\vec A|^2 \rangle} \int \frac{d^3 q}{(2 \pi)^3} \vec A( \vec k - \vec q) \cdot \vec A(\vec q) \, .
\end{equation}

If all the hidden photon modes are longitudinal, satisfying $\vec A(\vec k) = \hat k A_L(\vec k)$, then this becomes
\begin{equation}
\delta(\vec k) = \frac{1}{\langle A_L^2 \rangle} \int \frac{d^3 q}{(2 \pi)^3}  \frac{\vec q \cdot (\vec k - \vec q)}{q |\vec k - \vec q\,|}A_L( \vec k - \vec q) A_L(\vec q) \, .
\end{equation}

Since the primordial distribution for $A_L(\vec k)$ is gaussian, we can find the power spectrum of $\delta$ in the following way:
\begin{align}
\langle \delta (\vec k) \delta (\vec k') \rangle =& \frac{1}{{\langle A_L^2 \rangle}^2} \int \frac{d^3 q d^3 q'}{(2 \pi)^6}  \frac{\vec q \cdot (\vec k - \vec q)}{q |\vec k - \vec q\, |} \frac{\vec q\,' \cdot (\vec k' - \vec q\,')}{q' |\vec k' - \vec q\,'|} \times
\nonumber \\
&\left( \langle A_L( \vec k - \vec q) A_L( \vec k' - \vec q\,') \rangle \langle A_L(\vec q) A_L(\vec q\,') \rangle + 
\langle A_L( \vec k - \vec q) A_L(\vec q\,') \rangle \langle A_L(\vec q)A_L( \vec k' - \vec q\,') \rangle \right)
\\
=& (2 \pi)^3 \delta^3(\vec k-\vec k') \frac{8\pi^4}{{\langle A_L^2 \rangle}^2} \int \frac{q^2 d q \, d \phi \, d \cos \theta }{(2 \pi)^3} \frac{(\vec q \cdot (\vec k - \vec q))^2}{q^5 |\vec k - \vec q\, |^5} \mathcal P_A(|\vec k - \vec q\, |) \mathcal P_A(q) \, .
\end{align}
Integrating over $\phi$ and changing variables from $\cos \theta$ to $p = |\vec k - \vec q|$ gives the simple formula
\begin{equation}
\mathcal P_\delta (k) = \frac{k^2}{4 {\langle A_L^2 \rangle}^2} \int_{|q-k|<p<q+k}  \frac{(k^2 - q^2 - p^2)^2}{q^4 p^4} \mathcal P_A(p) \mathcal P_A(q) \, d q \, d p \, .
\label{eqn: density power spectrum}
\end{equation}
Using the results of section~\ref{sec:relic-abundance}, we calculate this for the vector field at late times. 
For small $k$, $\mathcal P_\delta$ falls as $k^3$, while at large $k$ it falls as $k^{-1}$.  The full numerically calculated power spectrum is shown in figure~\ref{fig:density-power-spectra}.

Note that the distribution of the density fluctuations are not gaussian. At long wavelengths, where the fluctuations about the mean are small, gaussianity is a good approximation. However at the special scale $k_*$, fluctuations are large and the distribution is non gaussian. On shorter length scales, the locally defined mean density varies dramatically from the global mean density, as can be see in Fig.~\ref{fig:field-profile}. On these scales it is not obvious that $\delta(\vec k)$ is even a useful quantity to consider, since it is defined globally.

\section{Careful treatment of the super-horizon regime}
\label{sec:vector-regime-careful}

In section~\ref{sec:evolution} we concluded that when $H$ is the largest scale in the equations of motion, the solution to the field equation is for $\A_L(\vec k, a)$ to remain constant. 
This is in spite of the fact that, for $k<k_*$, once modes enter the non-relativistic post-reheating regime, their equations of motion have a solution that grows linearly with $a$ until $a=a_*$. 
The reason this growing term can be dropped is that, prior to entering this regime, the equations of motion have already driven the $a$-derivative of $\A_L(\vec k, a)$ to be extremely small. 
We now check this fact more carefully.

We know that Eq.~(\ref{eq:long-horizon-crossing}) is the correct solution before and during horizon exit, up to possible corrections of order $m/H_I$ which could be present because $m=0$ was taken when deriving Eq.~(\ref{eq:long-horizon-crossing}). Shortly after horizon exit, Eq.~(\ref{eq:long-horizon-crossing}) behaves as
\begin{equation}
\qquad\qquad
\A_L(\vec k,a) \longrightarrow \A_0(\vec k) \bigg(1 + \frac{k^2}{2 a^2 H_I^2} + \frac{i k^3}{3 a^3 H_I^3} + \mathcal O\Big(\frac{k^4}{a^4 H_I^4}\Big) \bigg)
\qquad\qquad \text{(post-horizon exit).}
\end{equation}
We therefore write $\A_L(\vec k, a) = \A_0(\vec k)( 1 + \epsilon (k, a) )$, with $\epsilon$ behaving, shortly after horizon exit but still during inflation, as
\begin{align}
\qquad\qquad
\epsilon(k,a) &\longrightarrow \frac{k^2}{2 a^2 H_I^2}  + \frac{i k^3}{3 a^3 H_I^3}  + \mathcal O\Big(\frac{m}{H_I}\Big)
\qquad\qquad k/H_I \ll a \ll a_{\rm reheat} \, .
\label{eq:region-B-early-limit}
\end{align}
We then solve the equation of motion for $\epsilon$ and check that it remains small.

Using $\partial_t = a H(a) \partial_{a}$, the equation of motion for $\epsilon$ can be rewritten as
\begin{align}
\partial_a \Big(\frac{a^4 H(a)}{k^2+a^2m^2} \partial_a \, \epsilon \Big) &= -\frac{1}{H(a)} (1 + \epsilon) \approx -\frac{1}{H(a)} \, .
\label{eq:epsilon-eom}
\end{align}
$\epsilon$ has been dropped on the right hand side, since we are assuming it to be small and checking for consistency.
We know $H\approx H_I$ during inflation, and $H\approx m(a_*/a)^2$ as $a$ approaches $a_*$.  We can therefore approximate
\begin{equation}
\frac{1}{H(a)} = \frac{1}{H_I} + \frac{a^2}{a_*^2 m} \, .
\label{eq:H-approx}
\end{equation}
(It can be checked that deviations from this that occur well after horizon exit and well before $a=a_*$ do not affect the conclusion.)

The general solution to Eqs.~(\ref{eq:epsilon-eom}, \ref{eq:H-approx}) is
\begin{equation}
\begin{aligned}
\epsilon(k, a) 
&= c_1 + \frac{k^2}{2 a^2 H_I^2} + c_2 \frac{i k^3}{3 a^3 H_I^3} + \frac{m}{H_I} \bigg(i c_2 \Big( \frac{k^2}{k_*^2} + \frac{m}{H_I} \Big) \frac{k}{a H_I} + \frac{m}{H_I} \Big( \frac{4 k^2}{3 k_*^2} + \frac{m}{H_I} \Big) \ln\! \Big(\frac{k}{a H_I}\Big) \bigg)
\\&\qquad\qquad
 - c_2 \frac{i k m^2}{k_* H_I^2} \frac{a}{a_*} -  \Big( \frac{k^2}{6 k_*^2} + \frac{2m}{3H_I} \Big) \frac{a^2}{a_*^2} - \frac{1}{12} \frac{a^4}{a_*^4} \, .
\end{aligned}
\end{equation}
The terms in the first line dominate at the beginning of the period, and can be matched onto the desired behavior in Eq.~(\ref{eq:region-B-early-limit}), giving $c_1 = \mathcal O(m/H_I)$ and $c_2 = 1 + \mathcal O(m/H_I)$. 
The terms in the second line grow and dominate at the end of the period, near $a=a_*$. 
The first of these is the linearly growing term that we were concerned about. 
Having determined that $c_2 \approx 1$, we now see that this term is indeed incredibly suppressed.
Note that the remaining two terms, which are not suppressed by $m/H_I$, are simply the $O(1)$ corrections that are expected as the solution transitions into the matter-like regime at $a= a_*$.

\bibliography{Hidden-photon-DM}
\bibliographystyle{utphys}

\end{document}